\documentclass[iop]{emulateapj}

\usepackage{graphicx}
\usepackage{amssymb}
\usepackage{amsmath}
\usepackage{epstopdf}

\def \arcsec {\ensuremath{^{\prime\prime}}}
\def \psr {PSR\,J1023+0038}
\def \nustar {\emph{NuSTAR}}
\def \swift {\emph{Swift}}

\shorttitle{\nustar~ Observations of PSR\,J1023+0038}
\shortauthors{Tendulkar, S.~P. et al.}

\begin{document}

\setlength{\fboxsep}{0pt}%
\setlength{\fboxrule}{0.5pt}%

\title{\emph{NuSTAR} Observations of the State Transition of Millisecond Pulsar Binary PSR\,J1023+0038}

\author{Shriharsh P. Tendulkar\altaffilmark{1}, Chengwei Yang\altaffilmark{2,3}, Hongjun An\altaffilmark{2}, Victoria M. Kaspi\altaffilmark{2}, Anne M. Archibald\altaffilmark{4}, Cees Bassa\altaffilmark{4}, Eric Bellm\altaffilmark{1}, Slavko Bogdanov\altaffilmark{5}, Fiona A. Harrison\altaffilmark{1}, Jason W.~T. Hessels\altaffilmark{4,6}, Gemma H. Janssen\altaffilmark{4}, Andrew G. Lyne\altaffilmark{7}, Alessandro Patruno\altaffilmark{8,4}, Benjamin Stappers\altaffilmark{7}, Daniel Stern\altaffilmark{9}, John A. Tomsick\altaffilmark{10}, Steven E. Boggs\altaffilmark{10}, Deepto Chakrabarty\altaffilmark{11}, Finn E. Christensen\altaffilmark{12}, William W. Craig\altaffilmark{10,13}, Charles A. Hailey\altaffilmark{5}, William Zhang\altaffilmark{14}}

\email{spt@astro.caltech.edu}

\altaffiltext{1}{California Institute of Technology, 1200 E California Blvd, MC 249-17, Pasadena, CA 91125, USA}
\altaffiltext{2}{Department of Physics, McGill University, 3600 University St, Montreal, QC, Canada H3A 2T8}
\altaffiltext{3}{National Space Science Center, Chinese Academy of Sciences, 1 Nanertiao, Zhongguancun,Beijing,China 100190}
\altaffiltext{4}{ASTRON, The Netherlands Institute for Radio Astronomy, Postbus 2, 7990 AA, Dwingeloo, NL}
\altaffiltext{5}{Columbia Astrophysics Laboratory, Columbia University, 550 West 120th Street, New York, NY 10027, USA}
\altaffiltext{6}{Astronomical Institute `Anton Pannekoek', University of Amsterdam, Postbus 94249, 1090 GE Amsterdam, NL}
\altaffiltext{7}{Jodrell Bank Centre for Astrophysics, School of Physics and Astronomy, The University of Manchester, Manchester M13 9PL, UK}
\altaffiltext{8}{Leiden Observatory, Leiden University, PO Box 9513, NL-2300 RA Leiden, NL}
\altaffiltext{9}{Jet Propulsion Laboratory, California Institute of Technology, Pasadena, CA 91109, USA}
\altaffiltext{10}{Space Sciences Laboratory, University of California, Berkeley, CA 94720, USA}
\altaffiltext{11}{Kavli Institute for Astrophysics and Space Research, Massachusetts Institute of Technology, 70 Vassar Street, Cambridge, MA 02139, USA}
\altaffiltext{12}{DTU Space, National Space Institute, Technical University of Denmark, Elektrovej 327, DK-2800 Lyngby, Denmark}
\altaffiltext{13}{Lawrence Livermore National Laboratory, Livermore, CA 94550, USA}
\altaffiltext{14}{NASA Goddard Space Flight Center, Astrophysics Science Division, Code 662, Greenbelt, MD 20771, USA}


\begin{abstract}
We report \nustar\ observations of the millisecond pulsar - low mass X-ray binary (LMXB) transition system PSR\,J1023+0038 from June and October 2013, before and after the formation of an accretion disk around the neutron star. Between June 10--12, a few days to two weeks before the radio disappearance of the pulsar, the  3--79\,keV  X-ray spectrum was well fit by a simple power law with a photon index of $\Gamma=1.17^{+0.08}_{-0.07}$ (at 90\,\% confidence) with a 3--79\,keV luminosity of $7.4\pm0.4\times10^{32}\,\mathrm{erg\,s^{-1}}$. Significant orbital modulation was observed with a modulation fraction of $36\pm10$\,\%. During the October 19--21 observation, the spectrum is described by a softer power law ($\Gamma=1.66^{+0.06}_{-0.05}$) with an average luminosity of $5.8\pm0.2\times10^{33}\,\mathrm{erg\,s^{-1}}$ and a peak luminosity of $\approx1.2\times10^{34}\,\mathrm{erg\,s^{-1}}$ observed during a flare. No significant orbital modulation was detected. The spectral observations are consistent with previous and current multi-wavelength observations and show the hard X-ray power law extending to 79\,keV without a spectral break. Sharp edged, flat bottomed `dips' are observed with widths between 30--1000\,s and ingress and egress time-scales of 30--60\,s. No change in hardness ratio was observed during the dips. Consecutive dip separations are log-normal in distribution with a typical separation of approximately 400\,s. These dips are distinct from dipping activity observed in LMXBs. We compare and contrast these dips to observations of dips and state changes in the similar transition systems PSR\,J1824$-$2452I and XSS\,J1227.0$-$4859 and discuss possible interpretations based on the transitions in the inner disk. 

\end{abstract}

\maketitle

\section{Introduction}
Millisecond pulsars ~\citep[MSPs; ][]{backer1982} are neutron stars with surface magnetic fields  $B_\mathrm{surf}\sim10^{8}-10^{9}$\,G and rotation periods $P_{\mathrm{rot}}\lesssim30$\,ms that show radio, X-ray and/or $\gamma$-ray pulsations. The theory of recycled pulsars ~\citep{radhakrishnan1982, alpar1982, bhattacharya1991} suggests that during a low mass X-ray binary (LMXB) phase, angular momentum is transferred to the pulsar through disk-accretion from a binary companion. The consequent addition of angular momentum spins up the pulsar to high angular velocities. Further evolution disrupts the accretion and, as the ionized plasma in the pulsar magnetosphere diminishes, the pulsar may be observed as a radio MSP. In a few systems, the pulsar wind can ablate matter from its companion to form `black widow' \citep{fruchter1990} or `redback' \citep[see ][]{roberts2011} systems, sometimes leaving a planetary mass object~\citep{bailes2011}.

While the pulsar spin-up theory is well supported by the presence of binary companions around most MSPs and the discovery of accretion-induced millisecond X-ray pulsations in LMXBs~(\citealt{wijnands1998}; see \citealt{patruno2012} for a review) there is little understanding as to how and when the accretion stops and whether the transition from LMXB to a non-accreting MSP is swift and irreversible or whether the system flip-flops between the two states before settling into an non-accreting state~\citep{tauris2012}.  Recent observations of two remarkable LMXB-MSP transition systems, PSR\,J1023+0038~\citep{archibald2009} and PSR\,J1824$-$2452I~\citep{papitto2013}, were the first evidence of multiple state changes during the transformation.

The source FIRST\,J102347.6+003841 (later renamed \psr) was initially classified as a magnetic cataclysmic variable by \citet{bond2002}. \citet{thorstensen2005} suggested that the system was an LMXB before the confirming discovery of a 1.7-ms radio pulsar~\citep{archibald2009,archibald2010}. The pulsar is in a 4.75-hr orbit with a G type, $\sim$0.2\,M$_\odot$ companion. Double-peaked H and He lines in archival SDSS spectra revealed that the pulsar had an accretion disk during 2000--2001 but later spectra showed no evidence for accretion~\citep{wang2009}. This conclusion was supported by further optical and X-ray observations~\citep{archibald2010,bogdanov2011}. VLBI observations \citep{deller2012} of the pulsar allowed the measurement of its parallax distance ($1368^{+42}_{-39}\,$pc) and proper motion ($17.98\pm0.05\,\mathrm{mas\,yr^{-1}}$, $130\pm4\,\mathrm{km\,s^{-1}}$). Long term radio observations and $\gamma$-ray measurements have allowed the detailed understanding of the inclination, orientation and evolution of the system's orbit, the size and temperature of the companion star and estimates for the masses of both the components \citep{archibald2013}.

The pulsar has been monitored regularly and detected in the radio bands until recently, indicating the absence of accretion. In observations after June 15, 2013, the radio pulsations had decreased in flux to undetectable levels accompanied by a 20-fold increase in soft X-ray flux and a five-fold increase in the $\gamma$-ray flux~\citep[see ][]{stappers2013ATel,kong2013ATel,stappers2013,patruno2013}.

In this paper, we describe \nustar\ \citep{harrison2013} observations of \psr, the first hard X-ray (3--79\,keV) observations of this source. These observations were obtained during its quiescent state in June 2013, a few days before the pulsar's radio disappearance and later during the accretion phase in October 2013. This paper is organized as follows: Section~\ref{sec:obs} describes the details of the observations and X-ray data analysis. In Section~\ref{sec:results}, we describe the results of the spectroscopic fitting and timing analysis. The astrophysical implications of these results are discussed in Section~\ref{sec:discussion}.

\section{Observation and Analysis}
\label{sec:obs}
\psr\ was first observed by \nustar\footnote{\nustar\  is a 3--79\,keV focussing hard X-ray mission. It consists of two identical co-aligned Wolter-I telescopes with CdZnTe detectors at the focal planes. The telescopes provide a point-spread function with a full-width at half maximum (FWHM) of 18\arcsec and a half-power diameter (HPD) of 58\arcsec~ over a field of view of 12\arcmin$\times$12\arcmin. The energy resolution varies from 0.4\,keV at 6\,keV to 0.9\,keV at 60\,keV. The data from the two telescopes' focal plane modules are labelled FPMA and FPMB.} 
between June 10 and June 12, 2013 during a pre-scheduled $\approx$95-ks observation simultaneous with a $\approx$4-ks observation with the \swift\ X-ray Telescope~\citep[XRT; ][]{burrows2005}. The disappearance of \psr\ in radio monitoring  was constrained to have occured between June 15 and 30, 2013 \citep[see ][]{stappers2013} but the source could not immediately be re-observed by \nustar\ due to a solar angle constraint until mid-October. Based on the radio disappearance, a second 100-ks \nustar\ observation was scheduled simultaneous with a 10-ks \swift\ observation \citep[described in ][]{patruno2013} from October 19 to October 21, 2013. The details of the observations are summarized in Table~\ref{tab:obs}.

\begin{deluxetable}{lrrrr}
\centering
\tablecolumns{5} 
\tablecaption{Observations of \psr.\label{tab:obs}}
\tablewidth{0pt}
\tabletypesize{\footnotesize}
\tablehead{
  \colhead{Obs ID}   &
  \colhead{Start}      &
  \colhead{End}   &
  \colhead{Exp}      &
  \colhead{Rate}\\
  \colhead{}  &
  \colhead{(UT)}                    &
  \colhead{(UT)}                    &
  \colhead{(ks)}                 &
  \colhead{cts/s}
}
\startdata
\sidehead{\nustar}
30001027002 & Jun 10 13:15 & Jun 10 21:15 & 13 & 0.014\\
30001027003 & Jun 10 21:15 & Jun 11 14:50 & 34 & 0.011\\
30001027005 & Jun 12 05:35 & Jun 13 07:30 & 48 & 0.012\\
30001027006 & Oct 19 08:00 & Oct 21 17:45 & 100 & 0.4 \\
\sidehead{\swift\ XRT (Photon Counting Mode)}
00080035001 & Jun 10 14:03  & Jun 10 16:11 & 2.0 & 0.008\\
00080035002 & Jun 12 22:03  & Jun 13 00:00 & 1.9 & 0.011\\
00080035003 & Oct 18 05:08  & Oct 19 08:38 & 10 & 0.23\\
\enddata
\end{deluxetable}

The preliminary processing and filtering of the \nustar\ event data was performed with the standard \nustar\ pipeline version 1.2.0 and \emph{HEASOFT} version 6.14. The source was clearly detected at each epoch. We used the \texttt{barycorr} tool to correct the photon arrival times for the orbital motion of the satellite and the Earth. The source events were extracted within a 20\,pixel (49\arcsec, compared to a half-power-diameter of 58\arcsec) radius around the centroid and suitable background regions were used. Spectra were extracted using the \texttt{nuproducts} script. Using \texttt{grppha}, all photons below channel 35 (3\,keV) and above channel 1935 (79\,keV) were flagged as bad and all good photons were binned in energy to achieve a minimum of 30 photons per bin. 

Similarly, the \swift\ XRT data were processed with the standard \texttt{xrtpipeline} and the photon arrival times were corrected using \texttt{barycorr}. The \texttt{xrtproducts} script was used to extract spectra and lightcurves within a radius of 25\,pixels (59\arcsec). Photons in channels 0--29 (energy $<0.3$\,keV) were ignored and all channels between 0.3--10\,keV were binned to ensure a minimum of 30 photons per bin.

\section{Results}
\label{sec:results}

\begin{figure*}
\centering
\includegraphics[clip=true,trim=0in 0.0in 0.in 0.0in,width=0.95\textwidth]{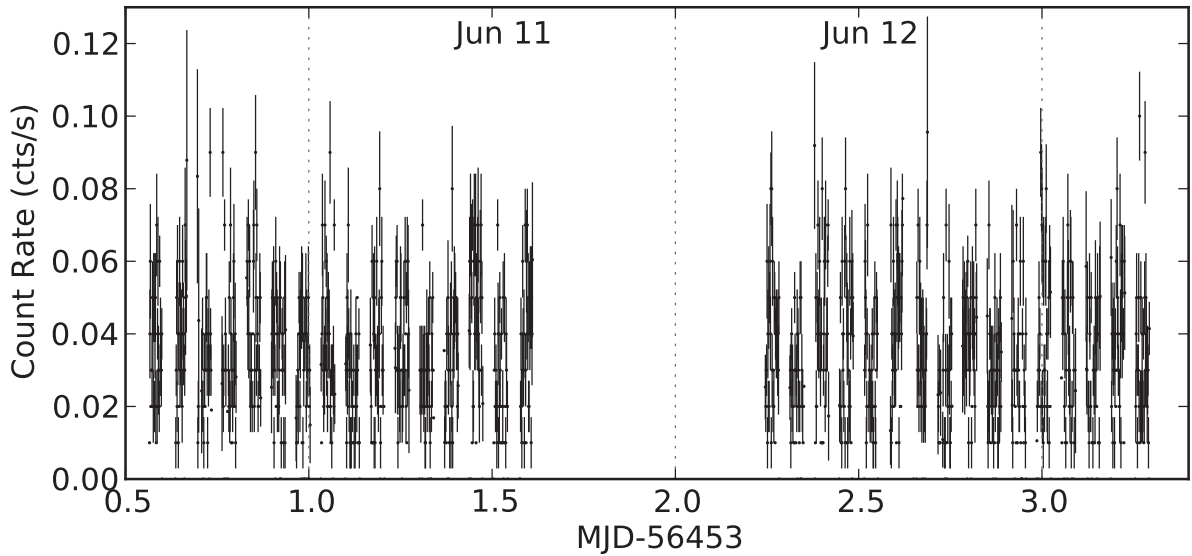}
\caption{100-s binned \nustar\ lightcurve of the June observations of \psr\ in the 3--79\,keV energy range. No flaring activity is observed within this observation window. The periodic gaps in the data are caused by Earth occultation of \nustar's line of sight to \psr\ through the spacecraft orbit. The dotted vertical black lines separate days of the observation.}
\label{fig:june_full_lc}
\end{figure*}

\subsection{June Observations}
The \nustar\ observation of \psr\ in June 2013 detected the source with an average count rate of $0.012\,\mathrm{cts\,s^{-1}}$ combined in FPMA and FPMB. No significant flaring or dipping activity was observed (Figure~\ref{fig:june_full_lc}). 

\subsubsection{Timing Analysis}
\begin{figure}
\centering
\includegraphics[clip=true,trim=0.05in 0.0in 0.55in 0.0in,width=0.48\textwidth]{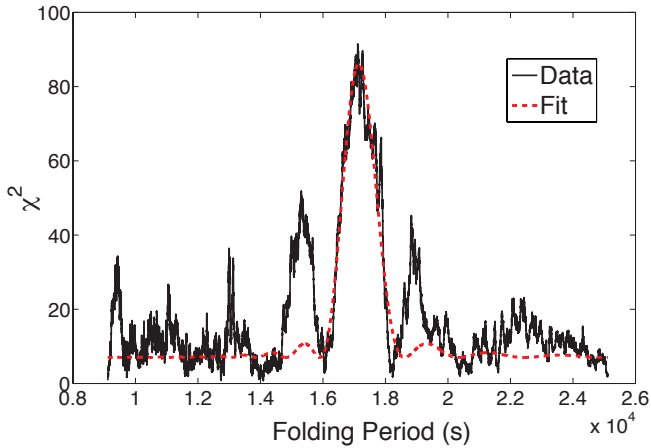}
\caption{$\chi^{2}$ test and epoch folding in searching for the binary-system orbital period with \nustar\ June data in the 3--79\,keV energy range. The solid black line is the $\chi^{2}$ of the folded observation data with respect to a null hypothesis as a function of folding period (i.e. test orbital period), while the red dashed line is the best fit of the data with the $\chi^2$ variation expected from a sinusoidal waveform \citep{leahy1987}. The $\chi^2$ distribution has 7 degrees of freedom.}
\label{fig:june_orbital_chisq_fit}
\end{figure}

We searched for orbital modulation by folding the \texttt{barycorr} corrected photons into 8 orbital phase bins with various trial orbital periods around the nominal period of 17100\,s (4.75\,hr). A range of orbital periods between 9000\,s and 25000\,s (2.5--7\,hr) were used with a step size of 1\,s. We used the epoch-folding statistic method~\citep{leahy1987} to fit the $\chi^{2}$ of the folded June observation data with respect to a null hypothesis. We detected the orbital modulation at a significance of $\approx$21-$\sigma$. After fitting (Figure~\ref{fig:june_orbital_chisq_fit}), the best-fit orbital period is 17148$\pm$83\,s. This is consistent with the  17115.52524(3)\,s  orbital period measured from long term radio monitoring \citep{archibald2013}. Varying the number of orbital phase bins and step sizes for orbital periods led to the same result.  

\begin{figure}
\centering
\includegraphics[clip=true,trim=0.4in 0.0in 0.55in 0.0in,width=0.48\textwidth]{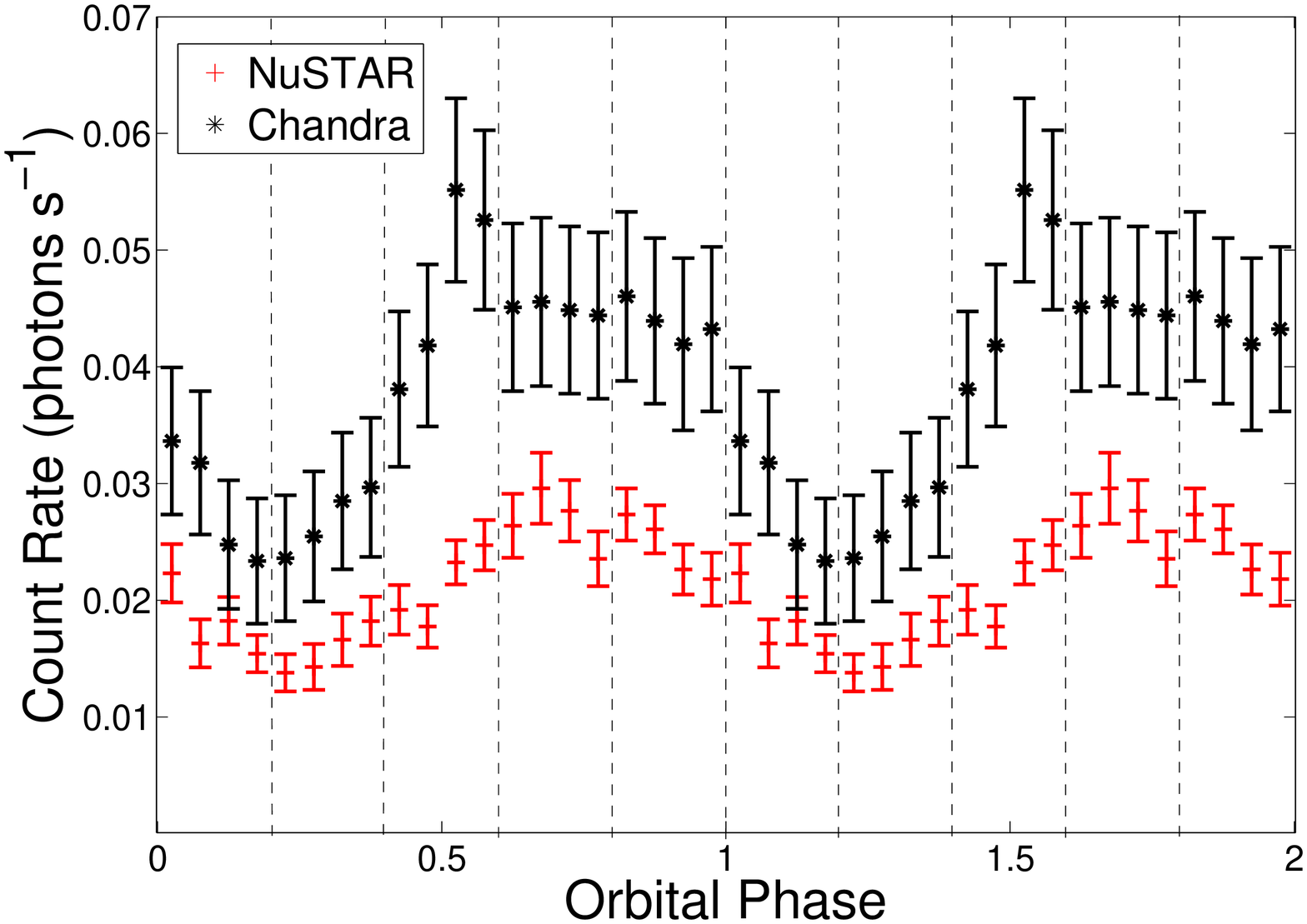}
\caption{Orbital modulation from \psr\ in the 3-79\,keV energy range in the June 2013 observation (red `+'). The corresponding 0.3-8\,keV \emph{Chandra} orbital modulation~\citep[black `*';][]{bogdanov2011} is overlaid. The \texttt{barycorr} corrected photons from FPMA and FPMB were folded with the best-fit period 17147.78\,s. The orbital modulation is plotted for two orbits for clarity.}
\label{fig:june_orbital_modulation}
\end{figure}

We folded the photons recorded in both FPMA and FPMB detectors with the best-fit period to create an orbital modulation profile (Figure~\ref{fig:june_orbital_modulation}). The modulation fraction, defined as $(F_{\mathrm{max}}-F_{\mathrm{min}})/(F_{\mathrm{max}}+F_{\mathrm{min}})$, where $F_{\mathrm{max}}$ and $F_{\mathrm{min}}$ are the maximum and minimum photon fluxes respectively, is $36\pm10$\,\%. The fractional root-mean-square modulation, defined as $\frac{\sqrt{\langle F^2\rangle-\langle F\rangle^2}}{\langle F\rangle}$, where the average is taken over all orbital phase bins, is 22\,\%. Compared with the amplitude of $0.0317\pm0.0095\,\mathrm{cts\,s^{-1}}$ for the lower energy range observed by Chandra in 0.3-8\,keV \citep{bogdanov2011}, the amplitude with NuSTAR observation, which is $0.0158\pm0.0034\,\mathrm{cts\,s^{-1}}$ in 3-79\,keV, is smaller.

\label{sec:pulsation_search}
We searched for pulsations at the spin-period of the pulsar in the \nustar\ data in several different ways. First, we note that \nustar's onboard clock, which is corrected at every ground pass, has exhibited residual timing jumps of 1--2\,ms at unpredictable times on time-scales of days to months (Madsen et al., in prep); this makes a pulsation search at the 1.7-ms period of \psr\ problematic. Nevertheless, recognizing that there could, in principle, be some spans in which the clock is sufficiently stable to detect bright pulsations and that a short-term pulsation search may be more likely to succeed than a long-term search, we proceeded to search as described below. 

In all cases, we used the {\tt PREPFOLD} facility within the {\tt PRESTO} suite of pulsar search software to fold the data set into 20 phase bins using spin and orbital parameters nominally determined from radio timing pre-June 2013 \citep[see ][]{archibald2013}.  Each search described below was done for three energy ranges: 3--79\,keV, 3--10\,keV and 10--79\,keV and the June and October data were analyzed separately. 

First, we folded each of the June and October data sets at the nominal ephemeris. Additionally, to maintain full sensitivity given the known orbital period variations \citep[see ][]{archibald2013}, we searched in $T0$ (the epoch of periastron) space by varying it $\pm 10$\,s around its nominal value in steps of 0.2\,s, folding at each trial value. The largest $T0$ error in this grid, 0.1\,s, corresponds to a 1.5\% error in rotational phase for our searches and is well below the uncertainties due to observed systematic timing variations.

We also searched for a pulse in smaller time spans:  we broke the June and October data into sequences of duration equal to the orbital period of the binary.  Each sequence was further divided into six equal parts.  We individually folded and searched each of these parts for pulsations.  Also, parts at the same orbital phase were combined in each of the June and October data sets separately, and the combinations searched. We repeated the same searches using 12 equal orbital phase sections. In total, we searched more than ten thousand parameter combinations, using reduced $\chi^2$ statistics. Accounting for the number of trials, we found no significant pulsations.


\subsubsection{Spectral Analysis}

\begin{figure}
\centering
\includegraphics[clip=true,trim=0.0in 0.0in 1.0in 0.9in,width=0.48\textwidth]{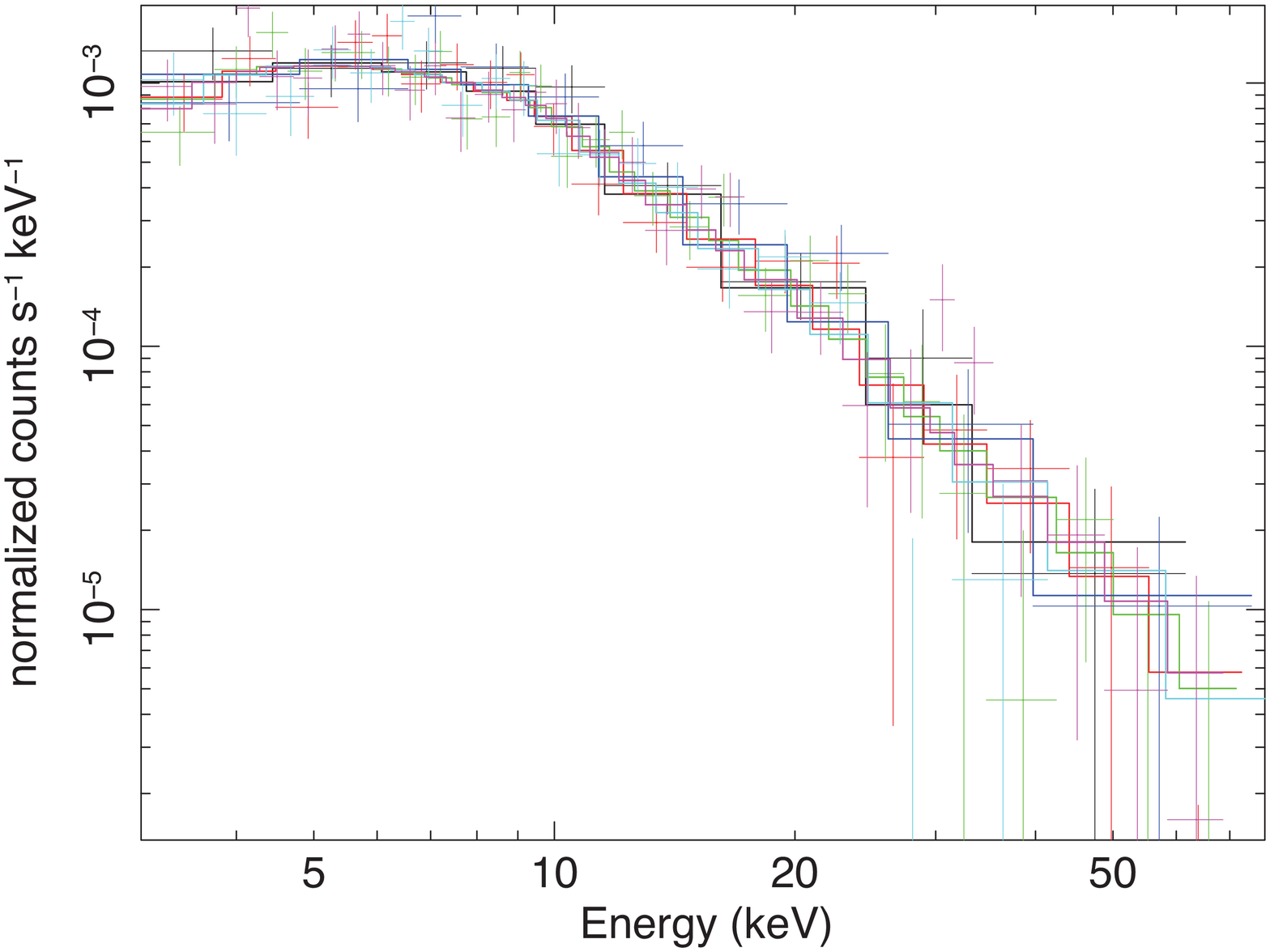}
\caption{Power law fit to June \nustar\ observations. The data sets \texttt{30001027002}, \texttt{30001027003}, and \texttt{30001027005} were fit simultaneously with the same model to improve signal-to-noise ratio. Column density was undetectable and hence set to zero. The data from the two \nustar\ detectors FPMA and FPMB were linked by a floating cross normalization constant. The fit achieved $\chi^2/\mathrm{dof}=113.4/117$. The correspondence between colors (in the electronic version of the manuscript) and spectra are as follows: black:\texttt{30001027002 FPMA}, red:\texttt{30001027003 FPMA}, green:\texttt{30001027005 FPMA}, blue:\texttt{30001027002 FPMB}, cyan:\texttt{30001027003 FPMB}, and magenta:\texttt{30001027005 FPMB}.}
\label{fig:june_average_spectra}
\end{figure}

The \swift\ XRT and \nustar\ spectra of \psr\ were fitted with an absorbed power-law model (\texttt{const*tbabs*powerlaw} in \texttt{XSPEC};  Figure~\ref{fig:june_average_spectra}). The source was barely detected in the \swift\ XRT observations \texttt{00080035001} and \texttt{00080035002} in June with 0.3--10\,keV count rates of $\lesssim10^{-2}\,\mathrm{cts\,s^{-1}}$. We added the two exposures to improve the signal-to-noise ratio. These observations were only used to constrain the estimate of the column density $N_\mathrm{H}$ during the June observation since \nustar\ data alone are not very sensitive to relatively low column densities. We constrained $N_\mathrm{H}<2.8\times10^{21}\,\mathrm{cm^{-2}}$ (3-$\sigma$), which is consistent with measurements by \citet{bogdanov2011} and \citet{archibald2010}. Setting $N_\mathrm{H}=0$ did not change the best-fit values of power-law index $\Gamma$ and the integrated flux; hence, $N_\mathrm{H}$ was frozen to zero for all future fits of the June data. 
No significant emission or absorption features are observed in the spectra. The thermal emission contribution observed by \citet{bogdanov2011} with $kT\approx$0.55--0.75\,keV is too faint in the 3--79\,keV band to be observed by \nustar. 

Table~\ref{tab:june_spectra} shows the measured values of $\Gamma$ and 3--79\,keV flux from the three June observations. The error bars are quoted at 90\% confidence. In subsequent analyses, the observations were simultaneously fit with a single model to improve the signal-to-noise ratio. The combined fit values are $\Gamma=1.17^{+0.08}_{-0.07}$ and $F_X=3.3\pm0.16\times10^{-12}\,\mathrm{erg\,cm^{-2}\,s^{-1}}$, corresponding to a 3--79\,keV luminosity of $7.4\pm0.4\times10^{32}\,\mathrm{erg\,s^{-1}}$ at \psr's measured distance. The fit achieved a $\chi^2$ of 113.4 with 117 degrees of freedom ($\mathrm{dof}$).

\begin{deluxetable*}{lcccc}
\tablecolumns{5} 
\tablecaption{\nustar\ Spectra During June 2013.\label{tab:june_spectra}}
\tablewidth{0pt}
\tabletypesize{\footnotesize}
\tablehead{
  \colhead{Parameter}   &
  \multicolumn{4}{c}{Observation (30001027xxx)}\\
\colhead{} &
  \colhead{002}  &
  \colhead{003}  &
  \colhead{005} &
  \colhead{Average}
}
\startdata
$\mathrm{C_{FPMB}}$\tablenotemark{a} &  $1.10_{-0.20}^{+0.23}$   &  $1.08_{-0.14}^{+0.16}$  &  $1.10_{-0.12}^{+0.13}$ & $1.087_{-0.085}^{+0.092}$   \\
$\Gamma$ ($N_\mathrm{H}=0$) & $1.00_{-0.17}^{+0.18}$ & $1.26_{-0.12}^{+0.13}$ & $1.15_{-0.10}^{+0.10}$ & $1.17_{-0.07}^{+0.08}$ \\
$\log_{10}(\mathrm{F_X})$\tablenotemark{b} & $-11.34_{-0.10}^{+0.10} $& $-11.55_{-0.08}^{+0.07}$ & $-11.47_{-0.07}^{+0.06}$ & $-11.48_{-0.05}^{+0.05}$  \\
$\chi^2/$dof & 10.9/16 & 46.57/41 & 48.14/57 & 113.4/117 \\
\enddata
\tablenotetext{a}{Scaling constant for FPMB data as compared to FPMA data.}
\tablenotetext{b}{3--79\,keV flux in units of $\mathrm{erg\,cm^{-2}\,s^{-1}.}$}
\end{deluxetable*}

\subsubsection{Orbital Modulation of Spectra}

\begin{deluxetable*}{lccccc}
\centering
\tablecolumns{6} 
\tablecaption{Orbital Variation of Spectral Fits in June\tablenotemark{a}.\label{tab:orb_mod}}
\tablewidth{0pt}
\tabletypesize{\footnotesize}
\tablehead{
  \colhead{Parameter}   &
  \multicolumn{5}{c}{Orbital Phase}\\
\colhead{} &
  \colhead{0.0-0.2}  &
  \colhead{0.2-0.4}  &
  \colhead{0.4-0.6}  &
  \colhead{0.6-0.8}  &
  \colhead{0.8-1.0}  
}
\startdata
%
PL Index ($N_\mathrm{H}=0$) & $0.97_{-0.23}^{+0.23}$ & $1.05_{-0.23}^{+0.24}$ & $1.29_{-0.17}^{+0.18}$ &  $0.96_{-0.16}^{+0.16}$ & $1.12_{-0.09}^{+0.09}$ \\
$\log_{10}(\mathrm{Flux})$\tablenotemark{b} & $-11.41_{-0.15}^{+0.14} $& $-11.59_{-0.14}^{+0.14}$ & $-11.53_{-0.10}^{+0.10}$ & $-11.25_{-0.10}^{+0.09}$  & $-11.47_{-0.05}^{+0.05}$ \\
\enddata
\tablenotetext{a}{Spectra from observations 2,3 and 5 were combined. Column density was undetectable and hence set to zero. The fit achieved $\chi^2/\mathrm{dof}=127.99/150$.}
\tablenotetext{b}{3--79\,keV flux in units of $\mathrm{erg\,cm^{-2}\,s^{-1}.}$ }

\end{deluxetable*}

\begin{figure}
\centering
\includegraphics[clip=true,trim=0.4in 0.0in 0.55in 0.0in,width=0.48\textwidth]{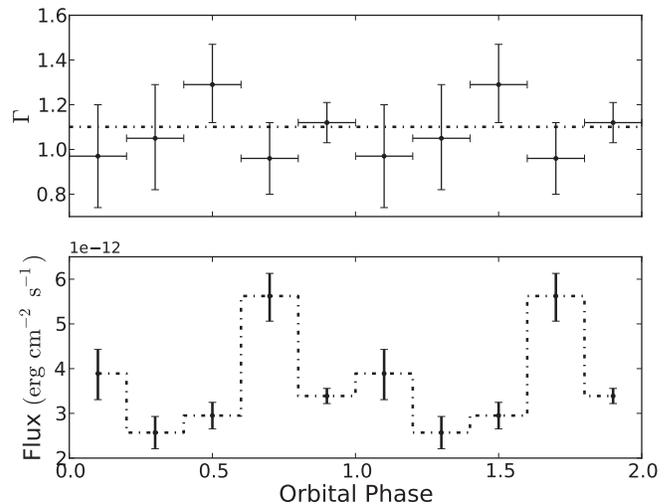}
\caption{\emph{Top Panel:}  Variation of power-law index ($\Gamma$) as a function of orbital phase for the combined June spectrum. Two orbits are shown for clarity. Black dots indicate best-fit values. All error bars are 90\% confidence. The dash-dotted lines indicate the error-weighted average of $\Gamma=1.10$.  The integrated 3--79\,keV flux from June (\emph{Bottom Panel}) is depicted by black dots.}
\label{fig:orbital_phase_modulation}
\end{figure}

To analyse the spectral variations during the orbit, we set \emph{good-time-interval} \texttt{(GTI)} windows for orbital phases: 0.0--0.2, 0.2--0.4, 0.4--0.6, 0.6--0.8, 0.8--1.0. To improve the signal of the phase-resolved spectra from the June observation, we summed up the events from \nustar\ observations \texttt{30001027002}, \texttt{30001027003} and \texttt{30001027005}. The five phase-resolved spectra extracted were fitted with an absorbed power-law model. From the previous discussion, the absorption column value was frozen to $N_\mathrm{H}=0$. The power-law index $\Gamma$ and normalization were allowed to vary for each phase.  Table~\ref{tab:orb_mod} and Figure~\ref{fig:orbital_phase_modulation} list and plot the best-fit values for $\Gamma$ and the integrated X-ray fluxes measured for the five orbital phases. The errors are quoted at 90\% confidence. The measurements are consistent with a constant $\Gamma$ value over the orbital phase with an error-weighted average of $1.10\pm0.12$.
 


\subsubsection{Comparison with Archival Data}
We compare the flux measured in June to archival measurements by \citet{homer2006,archibald2010} and \citet{bogdanov2011} (Table~\ref{tab:june_flux_wrt_archival}) by extrapolating the \nustar\ power law to lower energies. From the archival spectral fits, we extract the flux expected from the non-thermal power law since the soft thermal component is negligible in the \nustar\ energy band. We find that the fluxes are consistent within the expected cross-normalization errors (Madsen et al., in prep). Along with the detection of the radio MSP, this strongly suggests that on June 10 and June 12, 2013, \psr\ was in the same state with no accretion disk observed since 2004. 

\begin{deluxetable*}{lccccc}
\tablecolumns{6} 
\tablecaption{June 2013 Spectra Compared to Archival Measurements.\label{tab:june_flux_wrt_archival}}
\tablewidth{0pt}
\tabletypesize{\footnotesize}
\tablehead{
  \colhead{Reference}   &
  \colhead{Inst./Band (keV)} &
  \colhead{$\Gamma$} &
  \multicolumn{2}{c}{Flux} \\
  \colhead{}   &
  \colhead{} &
  \colhead{} &
  \multicolumn{2}{c}{($\mathrm{10^{-13}\,erg\,cm^{-2}\,s^{-1}}$)} \\
  \colhead{} & 
  \colhead{} &
  \colhead{} &
  \colhead{Meas.\tablenotemark{a}} &
  \colhead{Extrap.\tablenotemark{b}}  
}
\startdata
\citet{homer2006} & \textit{XMM-Newton}/(0.01-10) & $1.27\pm0.03$ & $5.3\pm0.6$ & $6.3\pm0.5$ \\ 
\citet{archibald2010} & \textit{XMM-Newton}/(0.5-10) & $0.99\pm0.11$& $4.9\pm0.3$ & $5.8\pm0.4$ \\ 
\citet{bogdanov2011}  & \textit{Chandra}/(0.3-8) & $1.00\pm0.08$ & $4.0\pm0.14$ & $4.9\pm0.4$ \\
\enddata
\tablenotetext{a}{Measured flux from only the non-thermal emission after separating the thermal soft-X-ray component, if any.}
\tablenotetext{b}{Extrapolated into the instrument's band from measured \nustar\ power law.}
\tablecomments{The flux errors do not include the systematic cross-normalization error.}
\end{deluxetable*}

\subsection{October Observations}

\begin{figure*}
\centering
\includegraphics[clip=true,trim=0.0in 0.0in 0.0in 0.0in,width=0.95\textwidth]{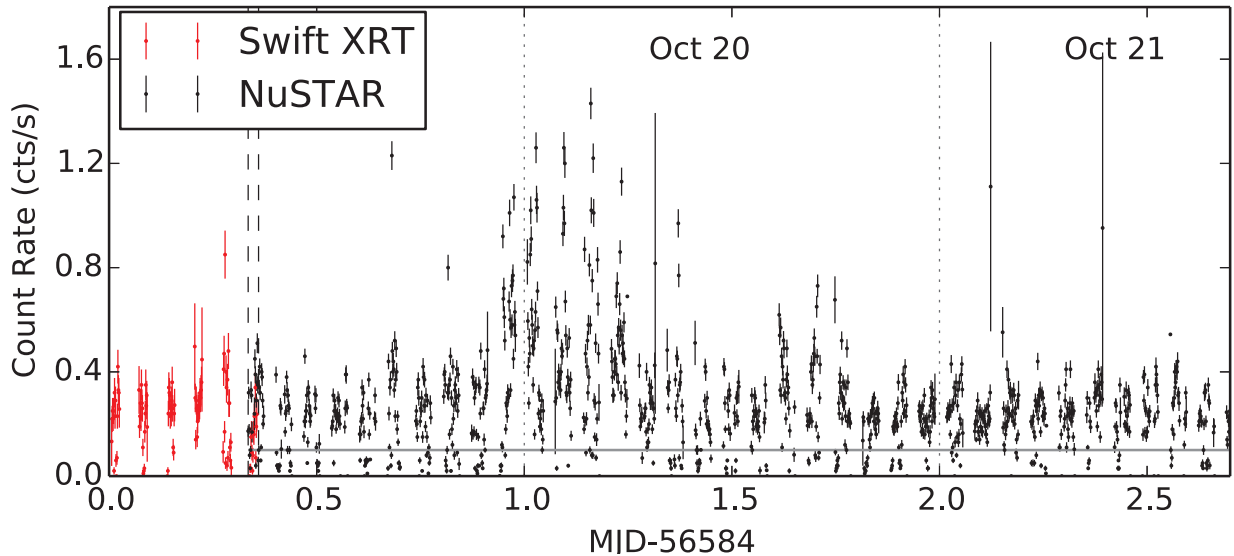}
\caption{100-s binned light curve from 0.3--10\,keV from \swift\ (red points) and 3--79\,keV from \nustar\ (black points) observations. For clarity, \nustar\ FPMA and FPMB observations are averaged. The average count rate during the $\approx$10\,hr long flare is factor of 5 higher than the out-of-flare count rate. The gray horizontal line at 0.1\,cts\,s$^{-1}$ denotes low flux levels seen in the \nustar\ observations discussed in detail in Section~\ref{sec:dips}. Two vertical dashed black lines denote the $\approx$30\,min overlap in the \swift\ and \nustar\ observations. Dotted black vertical lines divide the UTC dates of the observations, as noted at the top of the plot. The gaps in observations are due to Earth occultations.}
\label{fig:swift_nustar_full_lc}
\end{figure*}

Figure~\ref{fig:swift_nustar_full_lc} shows the \swift\ XRT and \nustar\ lightcurves during the October observations averaged over 100\,s bins. The two vertical dashed lines denote the $\approx$30\,min overlap in the \swift\ and \nustar\ observations.  During the \nustar\ observations, we observed a flare for a period of $\approx$10\,hrs with a factor of five increase in X-ray flux. Binning the same lightcurve in 1200\,s bins, we observed factor of 25 variations in the count-rate within 1\,hr (3 bin points). In the analysis that follows, we analyze the data with and without excision of the flare data.

\subsubsection{Timing Analysis}

\begin{figure}
\centering
\includegraphics[clip=true,trim=0.1in 0.0in 0.55in 0.0in,width=0.48\textwidth]{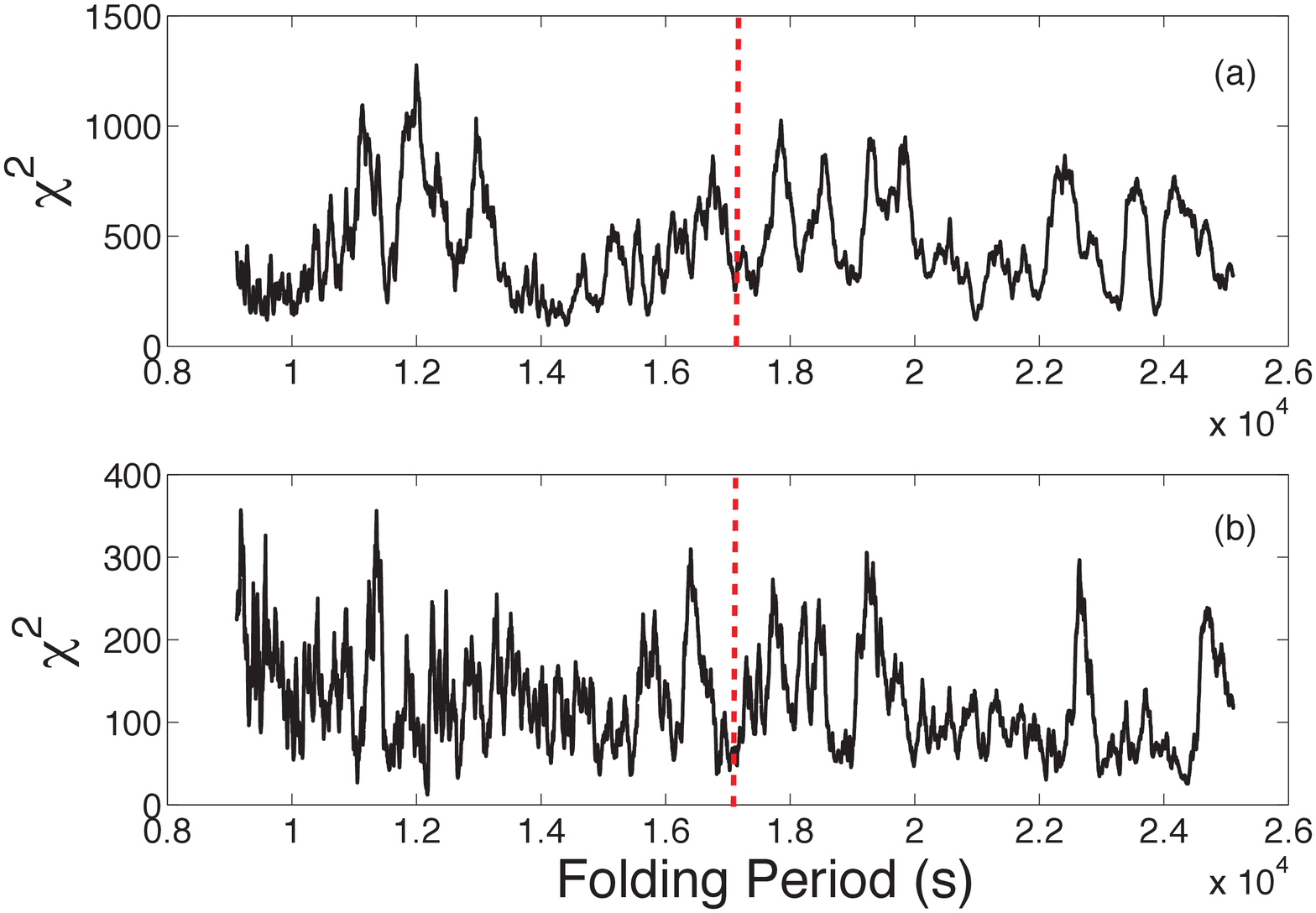}
\caption{\emph{Top Panel}: $\chi^{2}$ test and epoch folding in searching for the binary-system orbital period with \nustar\ October data in the 3--79\,keV energy range. The analysis is the same as done for the June data in Figure~\ref{fig:june_orbital_chisq_fit}. The solid black line is the $\chi^{2}$ of the folded observation data with respect to a null hypothesis as a function of trial orbital periods, while the red dashed line is the location of the peak measured in the June data.  The $\chi^2$ distribution has 7 degrees of freedom. \emph{Bottom Panel}: Same plot as above after excising the time periods with flares and dips. Although significant variation exists, mostly due to the flaring activity and variability, no clear signal of orbital period is observed.}
\label{fig:october_orbital_chisq_fit}
\end{figure}

\begin{figure}
\centering
\includegraphics[clip=true,trim=0.4in 0.0in 0.55in 0.0in,width=0.48\textwidth]{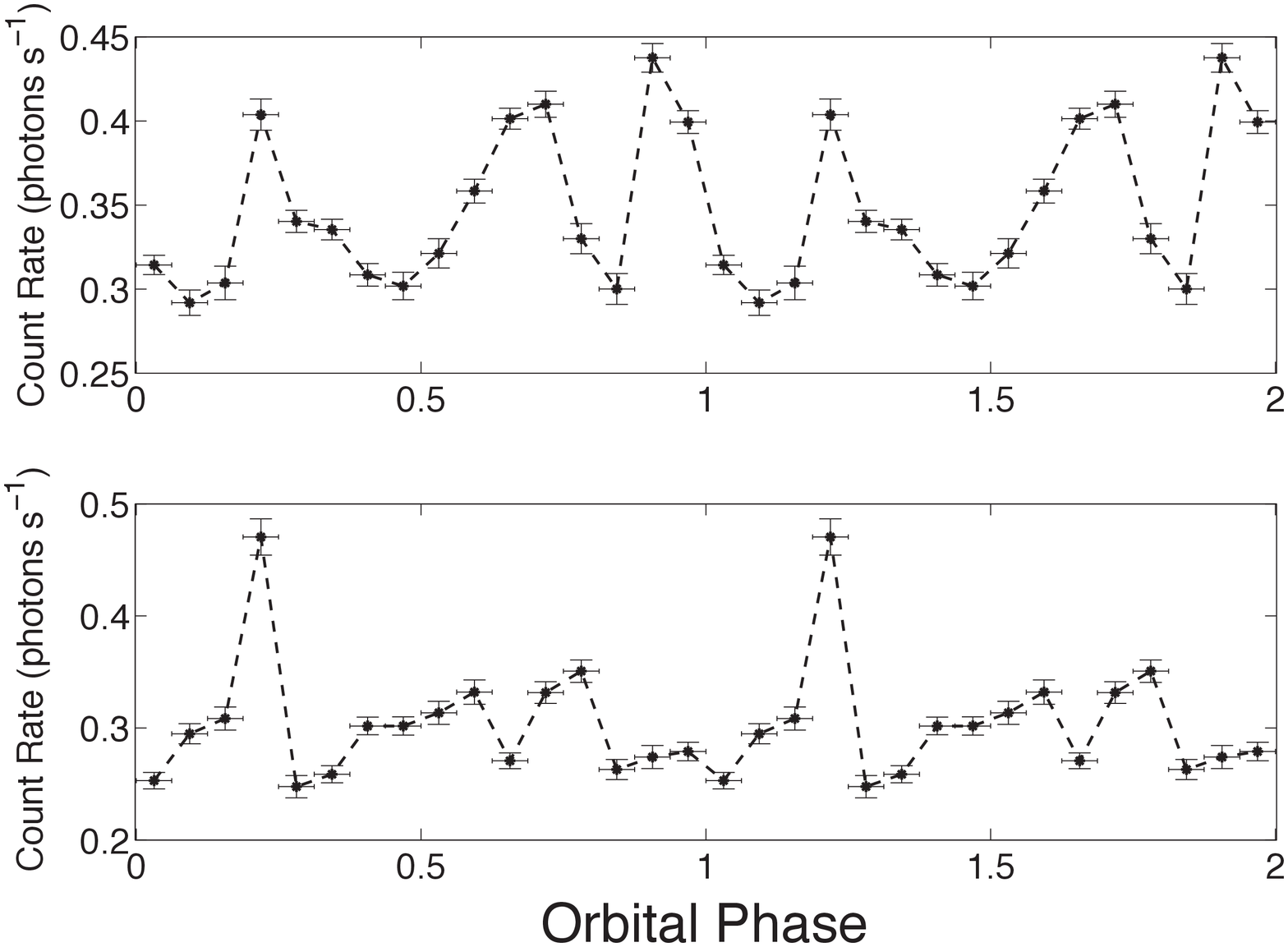}
\caption{\emph{Top Panel}: Light curve of October data in the 3--79\,keV energy range as a function of orbital phase, folded with the best-fit period of 17147.78\,s from June 2013. \emph{Bottom Panel}: Same plot after excising the time periods with flares and dips. No significant coherent sinusoidal modulation is observed. Although, significant variability is observed, it is not related to the orbit as such variability is observed regardless of folding period as is clear from the lack of a clear maxima in Figure~\ref{fig:october_orbital_chisq_fit}.}
\label{fig:october_orbital_modulation}
\end{figure}

We searched for the orbital modulation of X-rays from \psr\ using the same methodology as for the June data (Figures~\ref{fig:june_orbital_chisq_fit} and \ref{fig:june_orbital_modulation}). We folded the entire data set, as well as a subset for which flares and dips were excised, at a range of orbital periods centered around the value measured in June, however as shown in Figure~\ref{fig:october_orbital_chisq_fit}, practicallly all trial periods yielded a value of $\chi^2$ inconsistent with the null hypothesis.  This is strong contrast to what we observed in June (Figures~\ref{fig:june_orbital_chisq_fit} and \ref{fig:june_orbital_modulation}); in October, by contrast, we detect no evidence for orbital modulation. We folded the photons detected in both FPMA and FPMB detectors with the best-fit period of 17148\,s measured in June 2013 to create an orbital modulation profile (Figure~\ref{fig:october_orbital_modulation}) which also does not reveal coherent modulations, with or without the flare and dip data. We used a histogram of the measured $\chi^2$ values for the range searched and found the value of $\chi^2$ higher than 99\% of all other values. We then use the analytical formulae relating the amplitude and $\chi^2$ \citep{leahy1983} to derive an upper limit for the pulsed amplitude of $\approx$0.6\,cts\,s$^{-1}$ in the 3--79 keV band. However, we note that this estimate is only approximate, since the \citet{leahy1983} formulae assume that no significant power is present in the data, apart from the signal of interest, which is obviously not true in this case. A true upper limit is difficult to determine without more prior knowledge regarding the statistical nature of the strong variability observed.

\begin{figure}
\centering
\includegraphics[clip=true,trim=0.0in 0.0in 0.55in 0.0in,width=0.48\textwidth]{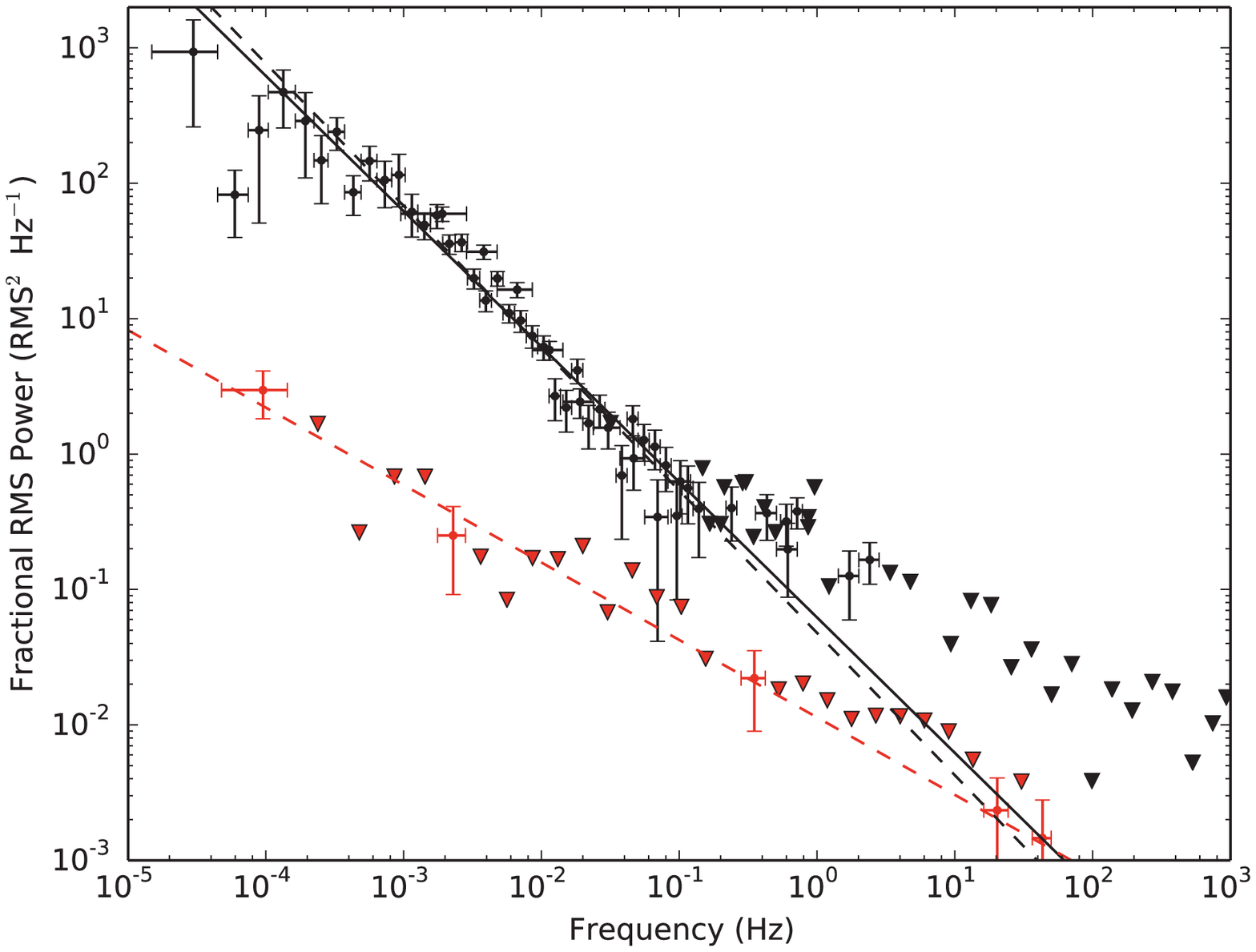}
\caption{Power spectral distribution (PSD) of \psr\ in October (black dots) normalized with the prescription of \citet{miyamoto1991}. The inverted black triangles indicate 3-$\sigma$ upper limits on the power spectrum. The PSD is well fit by a red noise (`flicker') power law, $P(f)\propto f^{-1}$ (solid black line). If the power law index is allowed to vary, the fit ($P(f)\propto f^{-1.05\pm0.05}$, dashed black line) is consistent with the flicker noise. Similarly, red dots and red inverted triangles denote the PSD during the June observations. The best-fit power law is $P(f)\propto f^{-0.57\pm0.03}$ (red dashed line).}
\label{fig:psr_powerspectrum}
\end{figure}

In order to look for other periodicities or quasi-periodic oscillations sometimes observed in other LMXBs, we binned the photon event time series from the source and the background regions in 500\,$\mu$s time bins and performed a Fourier transform to obtain the power spectrum of \psr. We used the fractional root-mean-square (RMS) normalization prescribed by \citet{miyamoto1991} such that the power spectrum is expressed in units of $\mathrm{(RMS/mean)^2\,Hz^{-1}}$. The expected Poisson rate was subtracted. We verified that the dead-time corrected lightcurves showed the expected Poisson variations at high frequencies in the background and source power spectra. Figure~\ref{fig:psr_powerspectrum} shows the power spectrum of \psr\ in October (black dots) and upper limits (black inverted triangles) compared to that in June (red dots and inverted triangles). We fit a red `flicker' noise power-law spectrum ($P(f)\propto f^{-1}$, black line) between $f=3\times10^{-4} - 5\times10^{-1}$\,Hz. While the standard $\chi^2$ value is 51 in 30 degrees of freedom, using the `whittle' statistic as discussed by \citet{barret2012} gives a value of 190 for 30 degrees of freedom. However, it is to be noted that the interpretation of these fits is not definitive for small number of degrees of freedom. The integrated fractional RMS variation in the above mention frequency band is 65\% and the residual fractional RMS variation, after subtracting the flicker noise component is 18\%. If the red noise power law index is allowed to vary, the best-fit power law is is $P(f)\propto f^{-1.05\pm0.05}$ (black dashed line), consistent with the `flicker' noise, however, the best-fit power law to the June power spectrum is $P(f)\propto f^{-0.57\pm0.03}$ (red dashed line). We do not see any significant features in the power spectrum.


Using the same procedures described in Section~\ref{sec:pulsation_search}, no X-ray pulsations at the pulsar spin period were detected in the October data.


\subsubsection{Spectral Analysis}

\begin{figure}
\centering
\includegraphics[clip=true,trim=0.25in 0.0in 1.0in 0.9in,width=0.48\textwidth]{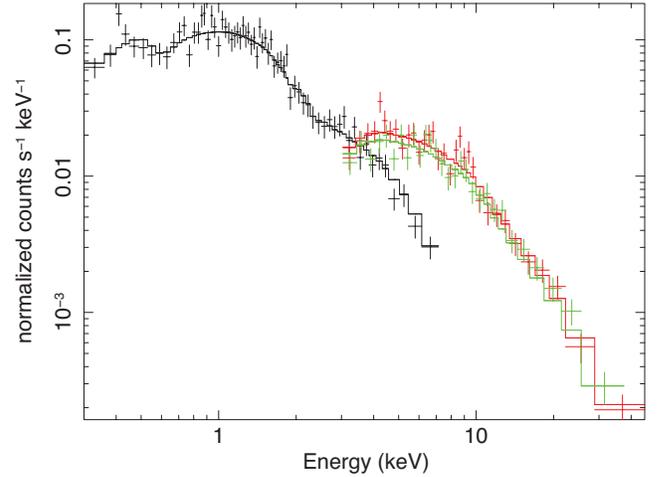}
\caption{Combined absorbed power-law fit to October observations with \swift\ XRT spectrum \texttt{00080035003} (black)  and \nustar\ spectrum \texttt{30001027006} (FPMA in red and FPMB in green). $N_\mathrm{H}$ was allowed to vary and achieved a best-fit value of $2.9_{-1.3}^{+1.5}\times10^{20}\,\mathrm{cm^{-2}}$ with $\Gamma=1.66_{-0.05}^{+0.06}$, consistent with \citet{patruno2013}. The fit achieved $\chi^2/\mathrm{dof}=120.55/124$.}
\label{fig:oct_swift_nustar_fit}
\end{figure}

The \swift\ XRT data and the first $\approx$4 hours of \nustar\ data were jointly fit (Figure~\ref{fig:oct_swift_nustar_fit}) with an absorbed power-law model. The cross normalization constants between \swift\ XRT and \nustar\ FPMA and FPMB modules were allowed to vary. The best fitting parameter values were: $N_\mathrm{H}=2.9_{-1.3}^{+1.5}\times10^{20}\,\mathrm{cm^{-2}}$ which is nominally consistent with the June spectra, $\Gamma=1.66_{-0.05}^{+0.06}$, with 3--79\,keV $F_X = 2.61\pm0.06\times10^{-11}\,\mathrm{erg\,cm^{-2}\,s^{-1}}$ and achieved a $\chi^2$/dof$=120.55/124$. The corresponding luminosity is $5.8\pm0.2\times10^{33}\,\mathrm{erg\,s^{-1}}$.  

\begin{figure}
\centering
\includegraphics[clip=true,trim=0.2in 0.0in 0.55in 0.0in,width=0.48\textwidth]{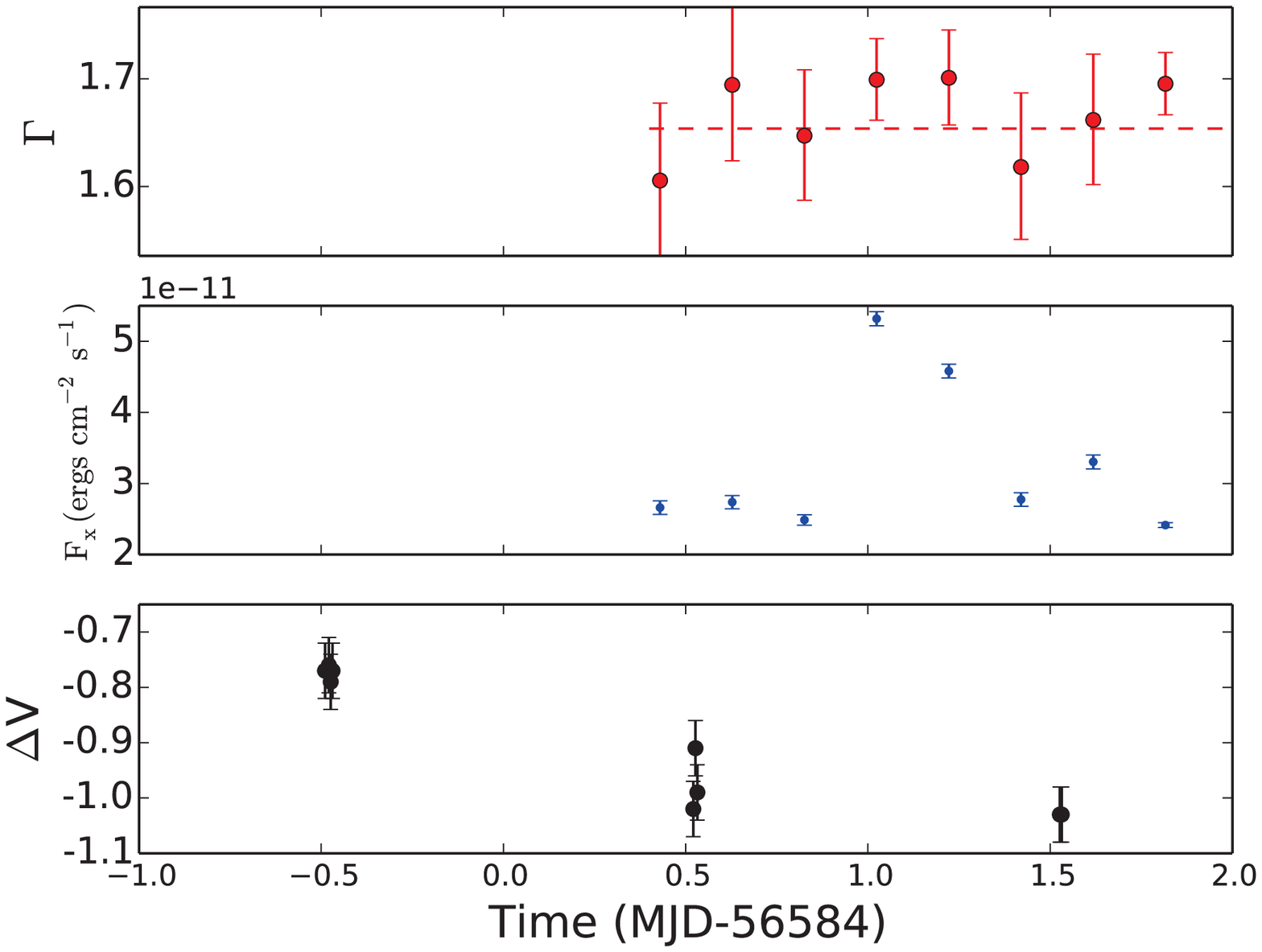}
\caption{\emph{Top Panel:} Variation of power-law index $\Gamma$ as a function of time during the \nustar\ October observation. Each data point is averaged over one orbital period to eliminate the effect of orbital modulation. The dashed line is the weighted average value of $\Gamma=1.654$. There is no statistically significant change in $\Gamma$ before, during or after the flare. \emph{Middle Panel:} Variation of 3--79\,keV X-ray flux measured by \nustar\ as a function of time. The flare is seen as a factor of two increase in flux at MJD$=$56585. \emph{Bottom Panel:} Corresponding variation in optical magnitude with respect to a neighboring star as a function of time. The data are from ~\citet{halpern2013ATel}. Due to the sparse monitoring, we do not have observations during the flare. However, the pre- and post-flare optical observations are consistent within the error bars. }
\label{fig:october_lightcurve_fit}
\end{figure}

In order to analyze the count-rate variation during the flare, we divided the \nustar\ October observation into 8 segments, each covering one orbital period of the system to eliminate any variation due to orbital modulation. We extracted spectra and fit an absorbed power-law to the measurements. We froze the value of the column density $N_\mathrm{H}=3.8\times10^{20}\,\mathrm{cm^{-2}}$ from the corresponding \swift\ observations~\citep{patruno2013}. Figure~\ref{fig:october_lightcurve_fit} shows the variation of the power-law index $\Gamma$ and X-ray flux as a function of time. The corresponding variation in optical magnitude  ~\citep{halpern2013ATel} is also plotted. Due to the sparse monitoring, there are no optical observations  during the flare. However, the pre-flare and post-flare optical brightness of the system is constrained to be almost equal. 

\subsubsection{Spectral Variations with Count Rate}
In order to analyze the variation in brightness, we created \texttt{GTI} windows by filtering the 100\,s lightcurve data points into six count-rate ranges: 0--0.1\,cts\,s$^{-1}$, 0.1--0.2\,cts\,s$^{-1}$, 0.2--0.3\,cts\,s$^{-1}$, 0.3--0.5\,cts\,s$^{-1}$, 0.5--0.7\,cts\,s$^{-1}$ and  0.7--2\,cts\,s$^{-1}$. We then re-extracted the spectra within the \texttt{GTIs} using the \nustar\ pipeline. The spectra are well fit by a PL model similar to the averaged spectra. The value of $N_\mathrm{H}$ was frozen to $3.8\times10^{20}\,\mathrm{cm^{-2}}$. 

\begin{figure}
\centering
\includegraphics[clip=true,trim=0.3in 0.0in 0.55in 0.0in,width=0.48\textwidth]{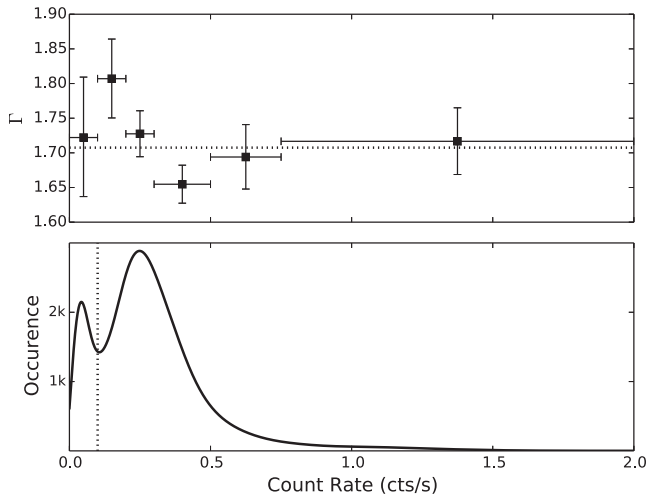}
\caption{ \emph{Top Panel:} Variation of photon PL index $\Gamma$ as a function of photon count rate. The photon count rates are binned between 0--0.1, 0.1--0.2, 0.2--0.3, 0.3--0.5, 0.5--0.7, 0.7--2.0\,$\mathrm{cts\,s^{-1}}$. $\Gamma$ varies only by 0.15 over a factor of 7 change in count rate. We detect a minor variation in the 0.1--0.2\,$\mathrm{cts\,s^{-1}}$ bin (transition between the two states) above a constant value (dotted line). \emph{Bottom Panel:} The distribution of count rates in the 100-s binned \nustar\ lightcurve (FPMA + FPMB) smoothed and weighted by the measurement errors. The dotted vertical line shows the demarcation between the two distinct flux states that are seen.}
\label{fig:october_countrate_analysis}
\end{figure}

Figure~\ref{fig:october_countrate_analysis} shows the variation of the 3--79\,keV flux and photon PL index $\Gamma$ as a function of count rate. There is weak evidence that the spectrum is harder at higher count rates, but the measurements of $\Gamma$ are consistent with a constant value over the four count rate ranges. The bottom panel of Figure~\ref{fig:october_countrate_analysis} shows a distribution of the count rates (weighted and smoothed by the measurement errors) from the 100-s binned lightcurve. There is clear evidence for two distinct states. The vertical dotted line shows the approximate count rate (0.1\,cts\,s$^{-1}$) which demarcates the two states. These two states occur due to sharp dips in the lightcurve, as described below. 


\subsubsection{Short Time-Scale Dips in the Light Curve}
\label{sec:dips}

\begin{figure}
\centering
\includegraphics[clip=true,trim=0.1in 0.0in 0.45in 0.3in,width=0.45\textwidth]{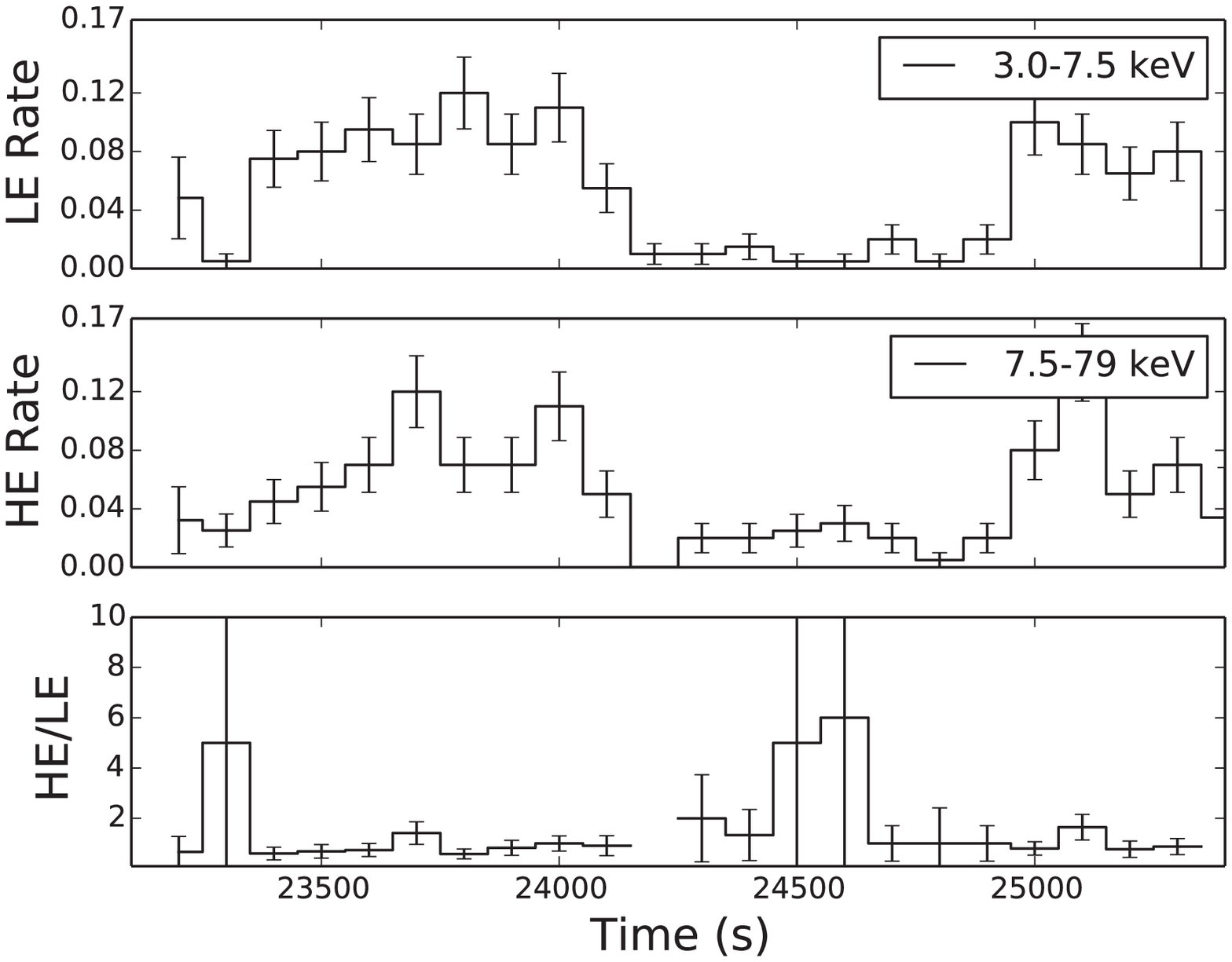}\\
\includegraphics[clip=true,trim=0.1in 0.0in 0.45in 0.3in,width=0.45\textwidth]{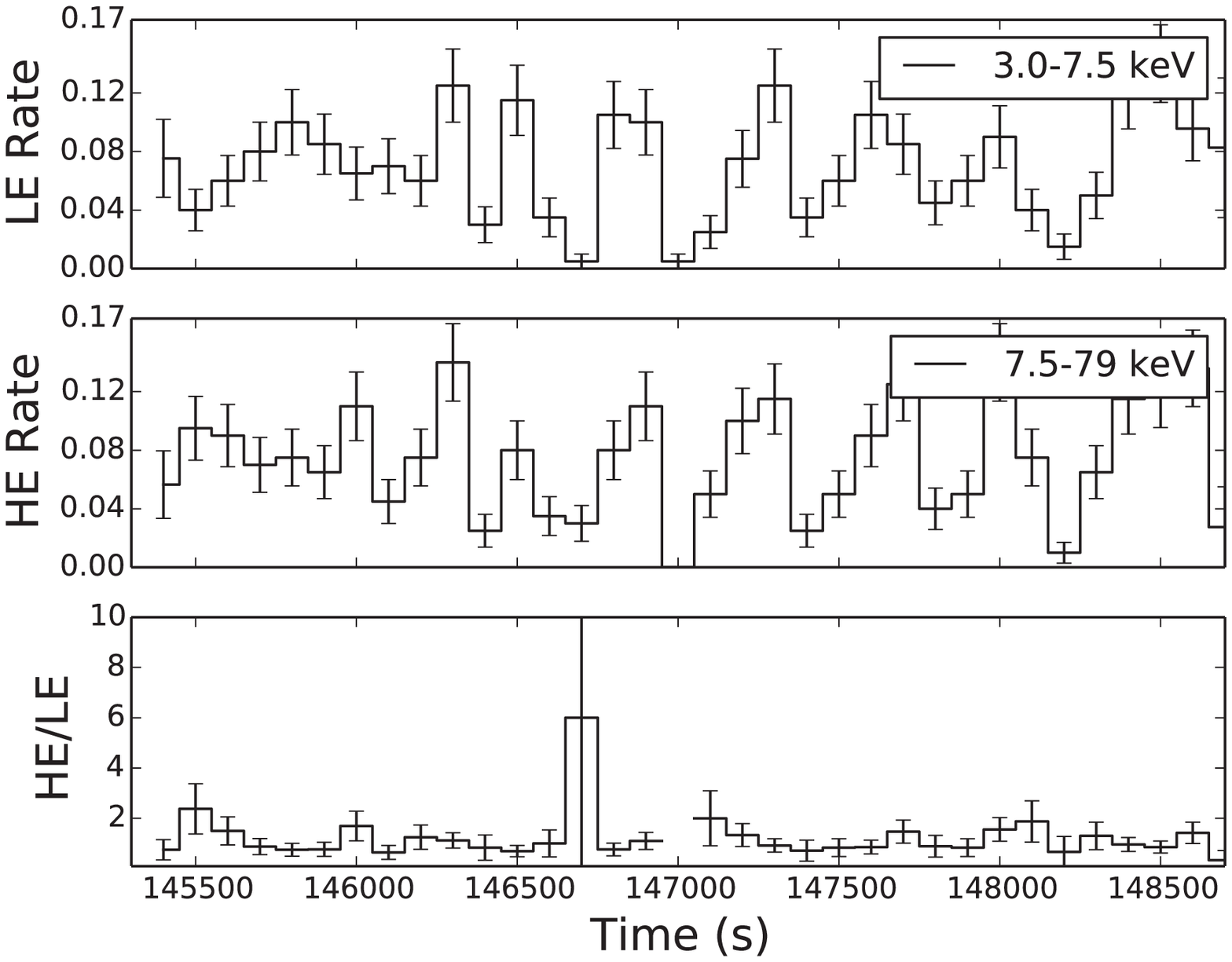}\\
\includegraphics[clip=true,trim=0.1in 0.0in 0.45in 0.3in,width=0.45\textwidth]{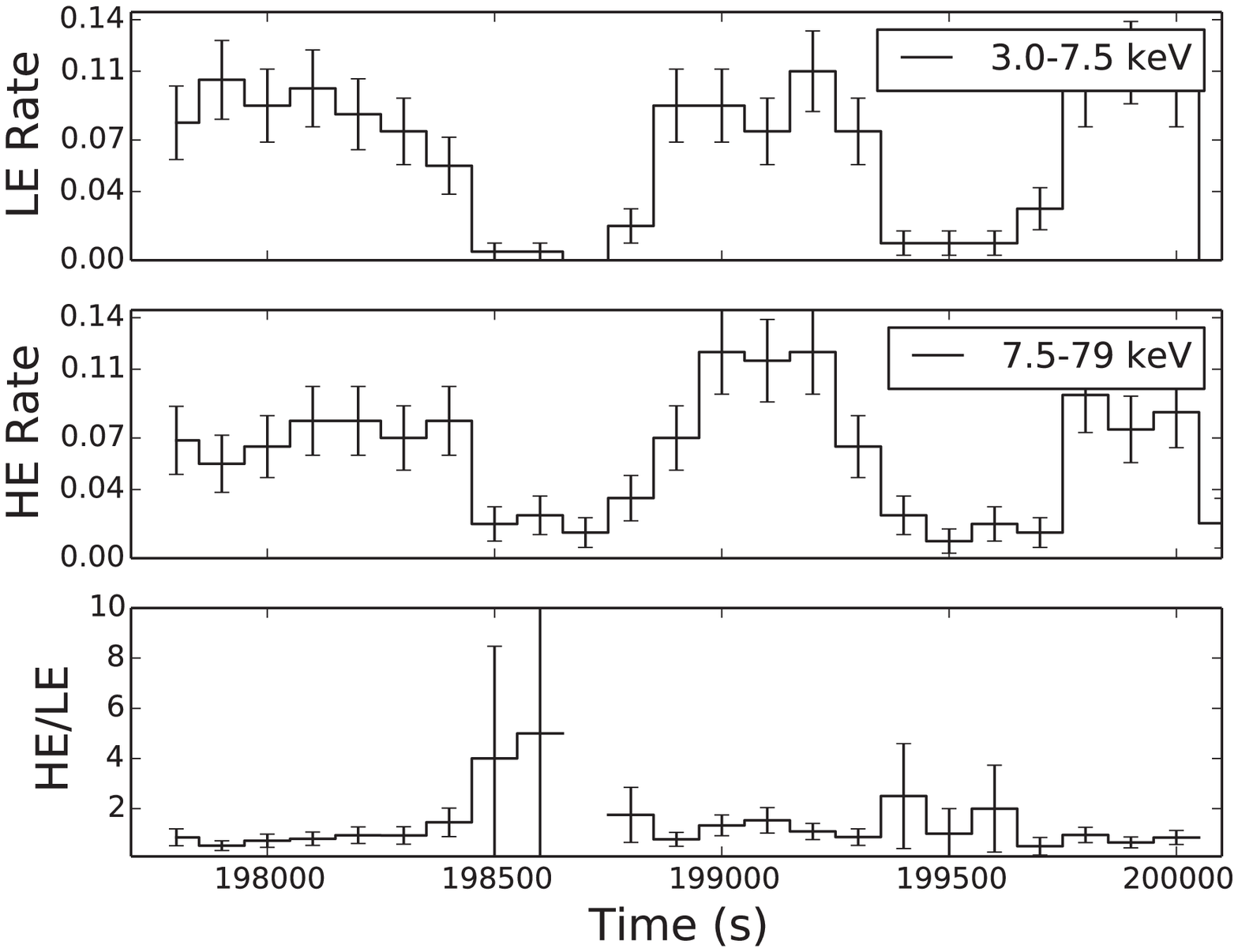}\\
\caption{Representative lightcurves from the October \nustar\ observation of \psr. In each figure, the top panel shows the low energy (3.0--7.5\,keV) count rate, the middle panel shows the high energy (7.5--79\,keV) count rate and the bottom panel shows the hardness ratio between the two energy bands. The hardness ratio is statistically constant unlike in dips observed in similar accreting systems. Each data point is averaged over 100-s time bins. The $x$-axis is time marked in seconds from the beginning of observation, 19 Oct 08:00:04 UTC.}
\label{fig:lightcurves}
\end{figure}

Apart from the 10-hr flare, significant short time-scale variations are observed during the October \nustar\ and \swift\ observations~\citep[see ][ for \swift\ XRT flickering]{patruno2013}. In this section, we present a detailed phenomenological description of the variations. Figure~\ref{fig:lightcurves} shows three examples of `dips' observed in the count rate binned in 100-s bins. The dips are not periodic or uniform in depth and width. In order to understand the nature of these variations, the photons were divided into low-energy (LE; 3--7.5\,keV) and high-energy (7.5--79\,keV)  bands.

In order to quantitatively analyze the occurence of these dips, we created a normalized lightcurve by dividing a 30-s binned lightcurve (of 3--79\,keV photons) with a lightcurve smoothed over long time-scale (600\,s). We verified that changing the smoothing time-scale between 600\,s and 1800\,s led to consistent results. The dips were defined as time periods when the normalized count-rate was less than 0.5. This threshold was determined by visual inspection of the normalized lightcurve and is insensitive to small variations (0.45--0.55). The dip periods thus extracted were smoothed with a binary closing function \citep[adapted from image processing; see ][ using the \texttt{scipy.ndimage} library]{gonzalez1992} to exclude small-amplitude, single-bin spikes in  wide ($>$4 bins) dips. The widths and positions of all other dips remain unaffected by this smoothing function. 

\begin{figure}
\centering
\includegraphics[clip=true,trim=0.4in 0.0in 0.55in 0.0in,width=0.48\textwidth]{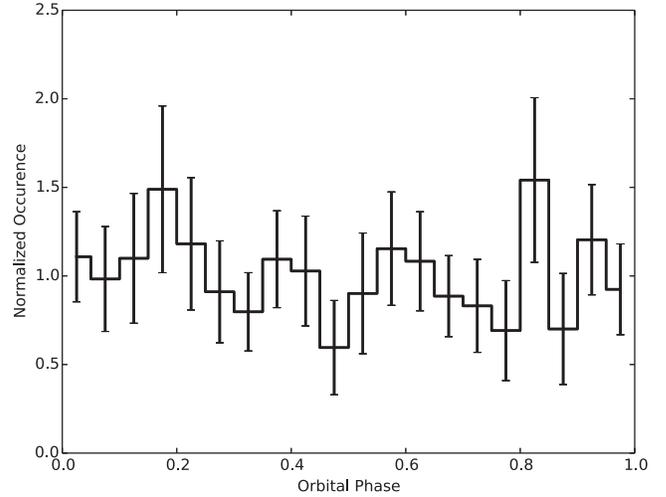}
\caption{Histogram of orbital phase distribution of dip centers. The distribution is normalized by the orbital phase coverage of the data which is uneven due to the near 1:3 ratio between the \nustar\ orbital period ($\approx$1.5\,hr) and the binary orbital period and the Earth occultations that interrupt observations every half revolution. The data are consistent with being uniformly distributed through orbital phases.}
\label{fig:dip_orbital_phase_histogram}
\end{figure}

We observed 224 dips in the entire observing sequence. To calculate the temporal positions of the centers of the dips in the normalized lightcurve, we converted the normalized lightcurve into a binary dip--non-dip lightcurve where the dip timepoints were marked as `1' and the non-dip timepoints were marked as `0'. The temporal positions of the dips were calculated by using a segmenting function to identify separate dips and using a center of mass function to calculate the central position of each dip segment. We converted the temporal positions into orbital phases using the orbital ephemeris described in the timing analysis\footnote{We used $T0= 55361.592856125$\,MJD and $P_\mathrm{orb}=17115.52238$\,s. The 10--20\,sec drifting of the $T0$ described in the timing analysis is too small to affect this analysis.}. We created a histogram of the observed dip position as a function of binary orbital phase (Figure~\ref{fig:dip_orbital_phase_histogram}). We divided each phase bin with the orbital coverage in that phase bin and the overall ratio was normalized to unity. Figure~\ref{fig:dip_orbital_phase_histogram} shows no significant preference in dip time for a specific orbital phase bin. The dips are thus uniformly and randomly distributed with orbital phase and and do not show any obvious relationship to the system's orbit. Similarly, a search for periodicity in the dip occurences did not reveal any significant signal.

\begin{figure}
\centering
\includegraphics[clip=true,trim=0.4in 0.0in 0.55in 0.0in,width=0.48\textwidth]{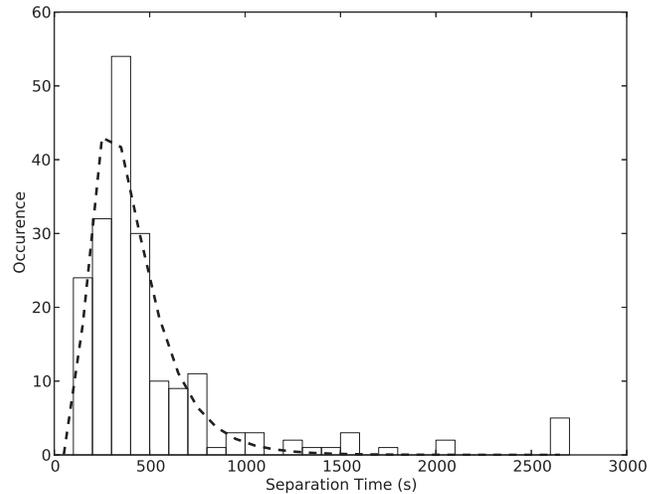}
\caption{Histogram of separations between the dip centers. The observations are well fit by a log-normal distribution with mean separation of $365\pm3$\,s and a scale factor of $0.48\pm0.04$ (see Section~\ref{sec:dips} for a description of the parameters). }
\label{fig:dip_separation_histogram}
\end{figure}

A histogram of the separations between consecutive dips is shown in Figure~\ref{fig:dip_separation_histogram}. The distribution is well fit by a log-normal distribution with a probability distribution function defined as,
\begin{equation}
P(x; A, \mu,\sigma)\mathrm{d}x = \frac{A}{x \sigma \sqrt{2 \pi}}\exp\left( -\frac{(\log x-\mu)^2}{2 \sigma^2}\right)\mathrm{d}x, 
\end{equation}
where $x$ is the separation of consecutive dips, $A$ is the normalization factor, $\mu$ is the location-parameter or the mean of $\log x$, and $\sigma$ is the scale factor of the distribution. We measure $\exp(\mu)=365\pm3$\,s and $\sigma=0.48\pm0.04$.

\begin{figure}
\centering
\includegraphics[clip=true,trim=0.4in 0.0in 0.55in 0.0in,width=0.48\textwidth]{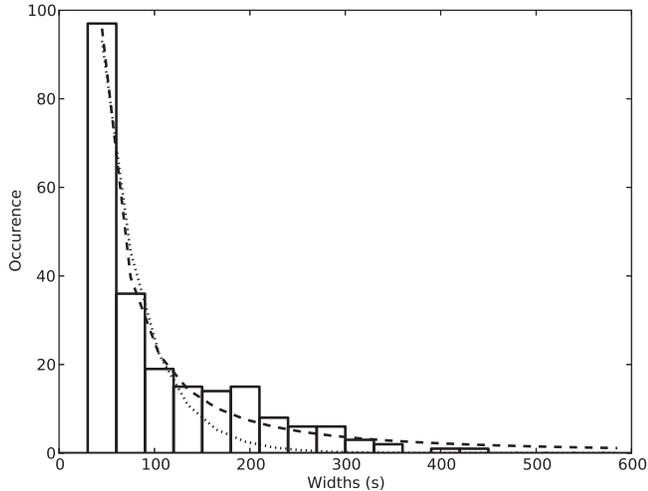}
\caption{Histogram of widths of the dips. The observations are fit by a power law (dashed line) with an index of $1.73\pm0.07$ with a reduced $\chi^2=1.3$ with 17 dof. An exponential law with a time-scale of $42\pm5$\,s (dotted line) may be fit, but the resulting $\chi^2$ is much worse (reduced $\chi^2\approx32$, with 17 dof) as it underestimates the occurence of dips with widths of 200--300\,s. The lowest time-scales have been excised from the fit to prevent statistical biasing from the smoothing algorithm described in the text.}
\label{fig:dip_width_histogram}
\end{figure}

Similarly, a histogram of the dip widths (Figure~\ref{fig:dip_width_histogram}) shows a sharp decline in the distribution as a function of dip width. The dips are well fit by a power law $P(x) \propto x^{-\alpha}$ where $x$ is the width of the dips in seconds. The best-fit value of $\alpha$ was measured to be $1.73\pm0.07$. The corresponding fit is shown as a dashed line in Figure~\ref{fig:dip_width_histogram}. An exponential fit to the distribution leads to a less significant fit (dotted line) with a best-fit time-scale of $42\pm5$\,s. Due to our smoothing function, we do not consider dips smaller than 60\,s in the fitting. The widths of the dips are not correlated with the separation to the preceding or successive dip.

\begin{figure}
\centering
\includegraphics[clip=true,trim=0.0in 0.0in 0.55in 0.0in,width=0.48\textwidth]{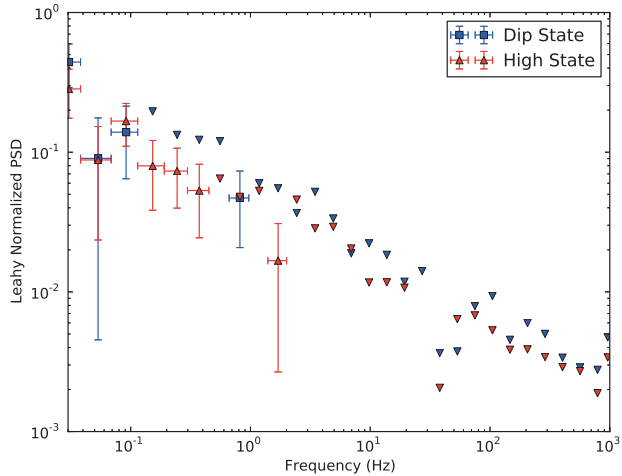}
\caption{PSD of \psr\ during dips (blue squares) and outside of dips (red triangles) normalized with the prescription of \citet{leahy1983}. Inverted blue and red triangles indicate 3-$\sigma$ upper limits for the PSD during dips and outside of the dips respectively. The two power spectra are statistically the same.}
\label{fig:dip_powerspectrum}
\end{figure}

We converted the dip locations into timing windows for the \texttt{xronos} task \texttt{powspec} and extracted the power spectrum during the dip states. We also created a power spectrum for the non-dip windows of the lightcurve. To avoid the variations caused by different Fourier windows, we chose contiguous intervals of 60-s length and averaged the powerspectra. Both the power spectra (Figure~\ref{fig:dip_powerspectrum}) were normalized using the normalization prescribed by \citet{leahy1983}. We do not observe any siginificant difference in the photon rate during the dip state as compared to the non-dip state.

\section{Discussion}
\label{sec:discussion}
We have presented 3--79\,keV observations of the MSP-LMXB transition system PSR\,J1023+0038 before and after the 2013 June 15-30 establishment of an accretion disk around the pulsar with phase-resolved spectral and timing analyses. We have observed the 3--79\,keV luminosity of the system increase from $7.4\times10^{32}\,\mathrm{erg\,s^{-1}}$ to $6.0\times10^{33}\,\mathrm{erg\,s^{-1}}$ between 2013 June 10-12 (few days to two weeks before the transition) and 2013 October 19-21. These luminosities, being much lower than the $10^{35-37}\,\mathrm{erg\,s^{-1}}$ luminosity typical of accreting LMXBs, are consistent with the previously proposed idea that Roche lobe overflow has occured, an accretion disk has formed, but the infalling matter has been prevented from accreting onto the pulsar surface and is now surrounding the pulsar~\citep[see][]{stappers2013ATel,kong2013ATel,halpern2013ATel,patruno2013,stappers2013}. 

\begin{figure}
\centering
\includegraphics[clip=true,trim=0.0in 0.0in 0.3in 0.0in,width=0.48\textwidth]{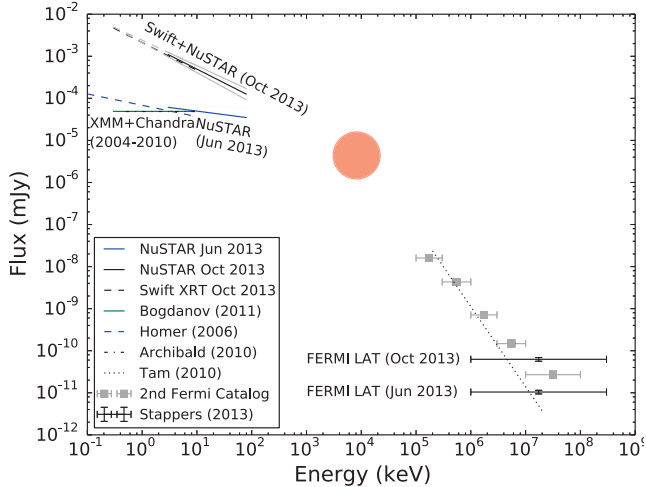}
\caption{A spectral energy distribution of \psr\ from published results and current work. The solid blue lines and the solid black lines in the 3--79\,keV band are the power-law fits from the June 2013 and October 2013 \nustar\ observations respectively. The light blue and gray solid lines denote the corresponding uncertainties at 90\,\% confidence. Black errorbars denote June and October flux estimates from \emph{Fermi}-LAT in the 1--300\,GeV band from \citet{stappers2013} which assumes $\Gamma=2.5$. The gray squares are photon fluxes from the \emph{Fermi}-LAT second source catalog~\citep{nolan2012}. The red circle is the approximate location of a possible crossover of the hard X-ray and $\gamma$-ray power-law: at a photon energy of about 1--10\,MeV.}
\label{fig:psr_SED}
\end{figure}

In order to perform a multi-wavelength comparison of the pre- and post-transition behavior, we compiled all published and new data into a spectral energy distribution (SED) from UV to $\gamma$-rays (Figure~\ref{fig:psr_SED}), augmenting the SED reported by \citet{takata2013}. The pre-transition (June) spectrum  from \nustar\  and the presence of deep orbital modulations is similar to the X-ray observations from 2004--2010~\citep{homer2006,archibald2010,tam2010,bogdanov2011} and is consistent with the hard X-rays being emitted by synchrotron emission from the intrabinary shock as suggested by \citet{bogdanov2011}. This similarity along with the sudden change in \emph{Fermi}-LAT $\gamma$-ray flux in June \citep{stappers2013} argues that the system in June was still in the diskless state achieved after its 2000-2001 accretion episode \citep{wang2009}. Hence, a gradual transition of the X-ray flux is ruled out, consistent with the radio behavior and $\gamma$-ray behavior \citep{stappers2013}.

From the presence of $\gamma$-ray emission and the absence of radio pulsations, \citet{stappers2013} suggested that the pulsar is enshrouded in gaseous material outside the `light-cylinder' radius of $81\,$km, held up by the magnetic pressure. Depending on the location of the barrier, the presence of material would either prevent the (a) detection or (b) generation of radio pulses. In the former case, the $\gamma$-rays could be emitted from the shock emission of the pulsar wind interacting with the infalling matter. In the latter case, the $\gamma$-ray emission may occur, in some  models, via shocks in a leptonic jet ejected by the propeller mechanism, as suggested by \citet{papitto2014} for the case of the LMXB-like state of XSS\,J1227.0$-$4859 which was recently shown to have undergone the reverse transition \citep{bassa2014} to a rotation-powered radio pulsar \citep{roy2014ATel}. The peak 3--79\,keV luminosity observed over the 4.75-hr orbital period is $1.2\times10^{34}\,\mathrm{erg\,s^{-1}}$. This is significant compared to the rotational energy loss of the pulsar $\dot{E}\approx4\times10^{34}\,\mathrm{erg\,s^{-1}}$, considering the low conversion efficiency between spin-down luminosity and X-ray luminosity seen in radio pulsars \citep[$\sim10^{-3}$ at this luminosity level;][]{possenti2002}.  The absence of orbital modulations in the post-transition lightcurve is consistent with the X-ray photons being emitted from near the transition region/corona at the inner edge of the accretion disk, possibly dominated by the synchrotron emission as suggested by \citet{papitto2014} for XSS\,J1227.0$-$4859 (with a very similar power-law spectrum with $\Gamma=1.7$). We observe that $\Gamma$ varies only by 0.05 over a factor of 7 change in count-rate during the flares, which is expected if the optical depth of Comptonization does not change significantly while the source photon population varies. The observed X-ray luminosity of \psr\ is in the range of $10^{-7}-10^{-6}\,L_{\mathrm{Edd}}$, far lower than for atoll sources \citep[which range from 0.01--0.5\,$L_{\mathrm{Edd}}$; e.g. see ][]{ford2000}. The low October X-ray luminosity compared with those of fully accreting neutron stars, along with the non-detection of pulsations, corroborates the absence of material accreting onto the surface of the neutron star. 

The lack of a high-energy cutoff in the \nustar\ energy range suggests that the highest electron energy in the pulsar wind is $\gtrsim79$\,keV. It is likely that the hard X-ray power law and the $\gamma$-ray power law crossover at photon energies of 1--10\,MeV (red circle in Figure~\ref{fig:psr_SED}). Observations of this energy range would add greatly to our understanding of this enigmatic system.

\subsection{Flat-Bottomed Dipping}
The flat-bottomed dips and flickering observed in the \nustar\ and \swift\ observations \citep{patruno2013} show the following characteristics that need to be explained by any theoretical interpretation.
\begin{enumerate}
\item{Luminosity variations from $\approx5\times10^{32}\,\mathrm{erg\,s^{-1}}$ in the low states to a nominal average level of $\approx6\times10^{33}\,\mathrm{erg\,s^{-1}}$,}
\item{Non-periodic occurence, uniformly distributed over orbital phase,}
\item{Nearly flat intensity in the bottom of the dip, with no significant hardness ratio change,}
\item{Ingress and egress time-scales between 10--60\,s,}
\item{No correlation between dip width and the separation to either the previous or next dip,}
\item{A log-normal distribution of dip separations as shown in Figure~\ref{fig:dip_separation_histogram} and}
\item{A decreasing distribution in dip widths as shown in Figure~\ref{fig:dip_width_histogram}.}
\end{enumerate}

This dipping activity is unlike activity observed in other similar systems, namely PSR\,J1824$-$2452I \citep{papitto2013b} and XSS\,J1227.0$-$4859~\citep{demartino2013}. In the 0.3--10\,keV range, \emph{XMM-Newton} observations of PSR\,J1824$-$2452I revealed sharp spectral changes as a function of count-rate \citep{ferrigno2013}. The ingress and egress time-scales of these dips were about 200\,s, the dip widths were up to a few thousand seconds and the low-state luminosity was $\approx10^{35}\,\mathrm{erg\,s^{-1}}$. The dips were interpreted as abrupt interruptions in accretion of matter onto the surface \citep[``weak'' and ``strong'' propeller regimes][]{illarionov1975,ustyugova2006}, reducing the X-ray luminosity and revealing partially obscured thermal emission from the neutron star surface. \citet{linares2013} reported that archival \emph{Chandra} observations of PSR\,J1824$-$2452I in quiescence (defined as $L_X/L_{\mathrm{Edd}}<10^{-4}$) revealed changes between ``active'' ($L_X$(0.3--10\,keV) $=3.9\times10^{33}\,\mathrm{erg\,s^{-1}}$) and passive ($5.6\times10^{32}\,\mathrm{erg\,s^{-1}}$) states with no change in the power-law spectral index ($\Gamma\approx1.5$). The transitions between the two states occurred on time-scales of 500\,s and the states lasted for $\sim10\,$h. The authors suggested that the state changes were caused by two different non-thermal emission mechanisms that coincidentally led to the same spectra: the high states being explained by magnetospheric accretion and the low states being attributed to intrabinary shock emission. Similar state change behavior observed in the quiescent LMXB EXO\,1745$-$28 in Terzan 5 was also attributed to variations in accretion rates ~\citep{degenaar2012}. 

\citet{demartino2013} observed dips with ingress and egress fast time-scales ($\lesssim10\,$s) in \emph{XMM-Newton} observations of  XSS\,J1227.0$-$4859 with dips widths between 200--800\,s. These dips were observed in the X-ray and near-UV bands but were absent in the ground-based optical observations, suggesting their origin close to the neutron star. The authors attributed dips occuring immediately after flares to sudden accretion onto the neutron star (`the flare'), the corresponding emptying (`the dip') and filling up of the inner regions of the accretion disk. Attempting to interpret the other dips as eclipses \citep[as in the case of LMXB dippers][]{church2004} with dense absorbing material covering a significant fraction of the X-ray source (i.e. the inner disk corona) did not lead to practical results, apart from the lack of periodicity. 

Recently, \citet{chakrabarty2014} reported highly variable `flickering' from Cen\,X-4 during recent observations where the \nustar-band luminosity was $\approx2\times10^{32}\,\mathrm{erg\,s^{-1}}$. While no `dips' or `eclipses' were observed during the observation, variability up to a factor of 20, on timescales of minutes to hours without significant spectral variability was observed. While Cen\,X-4 has not been directly observed to transition between an LMXB-like to a rotation-powered state, there may be some spectral evidence for such a change in archival data \citep{chakrabarty2014}. With the caveat of a small sample, it may be that such variability phenomena may be specific to the $10^{32}-10^{33}\,\mathrm{erg\,s^{-1}}$ luminosity range, where the systems transition between an LMXB-like and rotation powered MSP state.
 
In our data, the low luminosity (item 1, above), sets them apart from dips occuring due to interrupted accretion onto the neutron star surface. Item 2 makes our observations inconsistent with an association with the binary orbit or a specific radius in the accretion disk. Similarly, the orbital radii corresponding to the 10-s and 100-s time-scales are 7800\,km and 36000\,km, respectively, far larger than the expected location of hard X-ray source near the light cylinder radius, $r_{\mathrm{lc}}=81\,$km. Item 3 rules out eclipses due to optically thin material, suggesting instead either a change in the source brightness or an eclipse with a dense blob of material. However, the latter hypothesis is unlikely since the dense material would be expected to come to dynamical and thermal equilibrium with its surroundings in timescales of $t_{\mathrm{dynamical}} \sim \alpha t_{\mathrm{thermal}} \sim (H/R)^2 t_\mathrm{viscous}$ \citep*[see Eq. 5.68; ][]{frankkingraine2002}, where $\alpha$ is the Shakura-Sunyaev parametrization of disk viscosity \citep{shakura1973} and $H$ and $R$ are the disk height and radius, respectively. The viscous time-scale of the inner disk is estimated to be $\sim$10--100\,s \citep[see ][]{patruno2013} and as $H/R<1$ for an expected thin-disk, we expect these time-scales to be very short. 

Given the lack of change in the hardness ratio, it is possible that these variations are due to clumpy wind from the pulsar or clumpy accretion flow from the companion star. The viscous time-scale of the inner disk ($\sim$10--100\,s) is similar to the observed ingress and egress time-scales (Item 4). A possible explanation may involve the inner disk being pushed-back and reformed on these time-scales. The reduced mass transfer rate could diminish the source photon density available for Comptonization without significantly affecting the optical depth and electron temperature (and hence the power-law index $\Gamma$). However, it remains to be understood, (a), what mechanism would cause the disk to cycle through these states repeatedly; (b), why the ingress and egress times-scales are symmetrical and (c), why there is no correlation between the dip width and separation (Item 5). It would be interesting to understand how these changes occur at a fast time-scale in \psr\ as compared  to the long time-scale state changes observed in PSR\,J1824$-$2452I~\citep{ferrigno2013} and how they are affected by the accretion state and luminosity of the system. Items 6 and 7 are stated for completeness and need to be considered by a more detailed theoretical explanation but are outside the scope of this discussion.

Another candidate explanation invokes interrupted mass donations from the donor star that propagate through the accretion disk and are observed as dips. However, this scenario is unlikely because, (a), it is unlikely for the donor star to vary at 10--100\,s time-scale unless some seismological/tidal modes are active, in which case, periodicity would be expected; (b), the gaps propagating through the accretion disk would diffuse, leading to long ingress and egress time-scales, and (c), the gap propagation from the outer to inner disk would have been observed in different energies at different times, leading to varying hardness ratios.

The well-measured distance of \psr\ \citep[1.3\,kpc; ][]{deller2012} is much less than the typical 7--8\,kpc distances of other neutron star LMXBs~\citep{jonker2004},  making it a unique and important case study as it allows us to pursue detailed spectral and timing analysis at extremely low luminosity states which are unobservable for most other LMXB sources. Further monitoring of this source is warranted to reveal previously unobserved details of the transistions between LMXBs and MSPs.

\acknowledgements
We thank the anonymous referee for detailed suggestions and comments. This work was supported under NASA Contract No. NNG08FD60C, and made use of data from the {\it NuSTAR} mission, a project led by the California Institute of Technology, managed by the Jet Propulsion Laboratory, and funded by the National Aeronautics and Space Administration. We thank the {\it NuSTAR} Operations, Software and Calibration teams for support with the execution and analysis of these observations.  This research has made use of the {\it NuSTAR} Data Analysis Software (NuSTARDAS) jointly developed by the ASI Science Data Center (ASDC, Italy) and the California Institute of Technology (USA). VMK receives support from an NSERC Discovery Grant and Accelerator Supplement, from the Centre de Recherche en Astrophysique du Qu\'ebec, an R. Howard Webster Foundation Fellowship from the Canadian Institute for Advanced Study, the Canada Research Chairs Program and the Lorne Trottier Chair in Astrophysics and Cosmology. JWTH acknowledges funding for this work from ERC starting grant DRAGNET. AP acknowledges support from the Netherlands Organization for Scientiﬁc Research (NWO) Vidi fellowship.





\newpage

\bibliographystyle{apj}
\bibliography{paper}

\end{document}